\documentclass[12pt]{article} 
\usepackage[margin=1in,bmargin=1in,tmargin=1.2in]{geometry}
\usepackage{amsmath,amssymb,amsthm}
\usepackage{hyperref}
\usepackage{chngcntr}
\usepackage{physics}
\usepackage{lineno}
\usepackage{color}
\usepackage{float}
\usepackage{graphicx}
\usepackage{subcaption}
\usepackage{soul}
\usepackage{enumitem}
\usepackage{textcomp}  
\newtheorem{proposition}{Proposition}
\newcommand{\bydef}{\overset{\mathrm{def}}{=\joinrel=}}
\newcommand{\beq}{\begin{linenomath}\begin{equation*}} 
\newcommand{\eeq}{\end{equation*}\end{linenomath}} 
\newcommand{\beqn}{\begin{linenomath}\begin{equation}} 
\newcommand{\eeqn}{\end{equation}\end{linenomath}} 
\newtheorem{Def}{Definition}
\newtheorem{thm}{Theorem}

\newcommand{\R}{{\mathbb{R}}}
\definecolor{jim}{rgb}{.7,.4,0}

\definecolor{shayak}{rgb}{0,0.7,0.3}

\definecolor{sana}{rgb}{0.01,0.5, 0.1}

\newcommand{\aaa}{{a}}
\definecolor{rejector}{rgb}{0.69,0,0}

\newcommand{\black}{\color{black}{}}
 \newcommand{\red}{\color{red}{}}
 \newcommand{\blue}{\color{black}{}}
\definecolor{dgn}{rgb}{0,0.7, 0.2}

\definecolor{orange}{rgb}{1, .5, 0.0}

\newcommand{\err}{\text{RE}}
\newcommand{\ie}{\textit{i.e.}}
\newcommand{\BF}{\boldmath}

\providecommand{\keywords}[1]
{\small	
  \textbf{Keywords:} #1
}
\title{Ambiguity in the use of SIR models to fit epidemic incidence data}
 \author{B. Shayak, Sana Jahedi, and James A. Yorke}
\begin{document}
\maketitle
\date{}

\begin{abstract}  
When fitting a multi-parameter model to a data set, computer algorithms may suggest that a range of parameters provide equally reasonable fits, making the parameter estimation difficult. Here, we prove this fact for an SIR model. We say a set of parameter values is a good fit to outbreak data if the solution has the data's three most significant characteristics: the standard deviation, the mean time, and the total number of cases. In our model, in addition to the ``basic reproduction number'' $R0$,
three other parameters need to be estimated to fit a solution to outbreak data. We will show that those parameters can be chosen so that each gives a linear transformation of a solution's incidence data. As a result, we show that for {\it every} choice of $R0>1$, there is a good fit for each outbreak.

\black We also illustrate our results by providing the least square best fits of the New York City and London data sets of the Omicron variant of COVID-19. Furthermore, we show how versions of the SIR model with $N$ compartments have far more good fits- - indeed a high dimensional set of good fits -- for each target --  showing that more complicated models may have an even greater problem in overparametrizing outbreak characteristics. 

\black

\end{abstract}

\section{Introduction}
The mathematical modeling of epidemics dates back to at least the 1910s when Sir Ronald Ross constructed a simple differential equation to model the spread of the vector-borne disease malaria~\cite{ross1911prevention}. In 1927, 
William Kermack and Alexander McKendrick formulated a general mathematical theory of the spread of an epidemic~\cite{SIR}. 
As a special case of their general theory, they formulated an SIR model, a set of coupled first-order nonlinear differential equations describing the transmission of an infectious disease where infected persons are removed from the susceptible population upon recovery or death. 
They applied the model to an outbreak of plague in Mumbai, India. In recent years  SIR models, sometimes with slight variations, have been used for analyzing a variety of epidemics, including outbreaks of measles, chicken pox, mumps, polio~\cite{yorke1,yorke6,giraldo2008deterministic,mahmud2017comparative}, \black 
Ebola virus~\cite{ebola1,ebola2,ebola3}, Zika virus \cite{zika1,zika2,zika3}, Nipah virus~\cite{nipah1,nipah2,nipah3}, and many applications to {\blue COVID-19~\cite{COOPER2020110057,kudryashov2021,pillonetto2021tracking,BAGAL2020110154,yorke-jahedi,David2023}.}

The two main transmission parameters of an outbreak are $\tau$, the average duration of infectiousness of an infected individual, and 
 $R0$, the average number of individuals who are contacted (and become infected if they are susceptible) by an average infectious person. 
It is well known that it is difficult to determine $\tau$ and  $R0$ directly from incidence data~\cite{li2011failure,weitz2015,magal2018parameter}.  $R0$ has been estimated for different strains of COVID-19 using other types of data by $R0$ has been estimated for different strains of COVID-19 using other types of data by  consortia scientists, including those from The London School of Hygiene \& Tropical Medicine (LSHTM), Institute for Health Metrics and Evaluation (IHME), and Johns Hopkins University (JHU). 
Such measurements can be invaluable additions to results from fitting outbreaks. This paper emphasizes the impossibility of getting such parameter estimates by just fitting models to incidence data. 
We show the extent of that difficulty using theoretical and computational tools. We show that a wide variety of combinations of estimated parameters can fit the same outbreak almost equally well, confirming that it is very challenging to correctly estimate ``the basic reproduction number'' or the infectious period from case data. We specifically focus on one particular outbreak: that of B1.1.529, the ``Omicron'' variant of COVID-19 in winter 2021-2022.
In New York City, USA (NYC) and in London, UK, about 1 million infections were reported during this outbreak, a fraction of the actual Omicron infections.

\begin{figure}
    \centering
    \includegraphics[width=\linewidth]{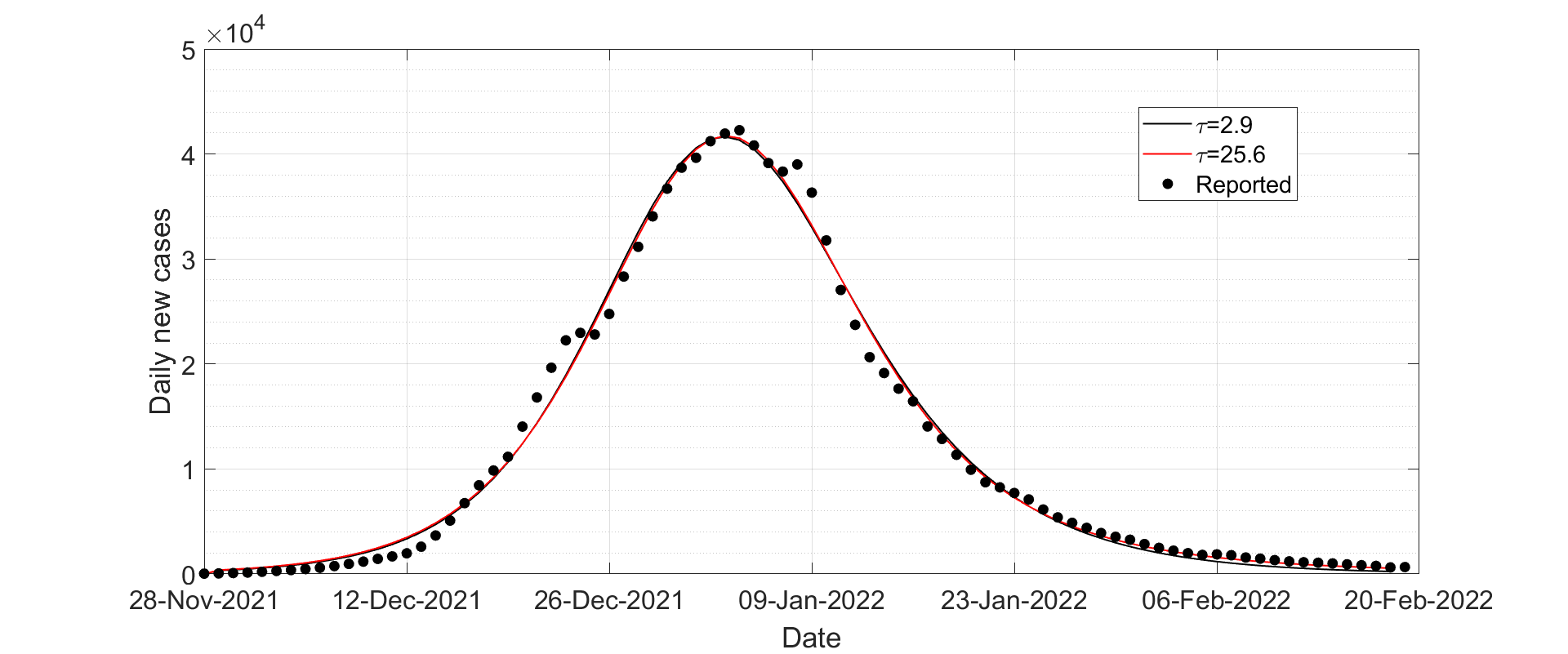}
    \caption{{\bf Fitting SIR solutions to NYC COVID-19 Omicron daily cases.} 
    The black dots show the daily new cases of Omicron in New York City from November 28, 2021 to February 20, 2022. 
      The solid black and red curves are the SIR solutions that are least-squares fits to the data for different values of the average duration of infectiousness, $\tau = 2.9$ and $25.6$ 
      (See Section~\ref{subsec:least squares}.)
      As the plot shows, both fits are almost equally good, and these curves are almost identical. The red curve shows the outbreak from the best least squares fit for all $\tau$ between 2 and 30, namely $\tau=25.6$
      More details will be shown later in Figure~\ref{NYC 2 parts}}. 
    \label{fig:NYC best fit}
\end{figure}

 We show two least-squares fits for the NYC COVID-19 Omicron data in Figure~\ref{fig:NYC best fit}, one in red and one in black. This picture starkly presents the challenge with fitting: two SIR least-squares fits to the NYC COVID-19 Omicron new case data are almost identical despite having quite different parameter values. See Table~\ref{suptable nyc} in Supplementary Material for the parameter values. 
 
 \black The two fits are so close to each other that they almost appear to be one curve, so the fits cannot be used to estimate the values of the parameters.

\begin{figure}%
\begin{subfigure}[c]{\columnwidth}
 \centering
    \includegraphics[width=\linewidth]{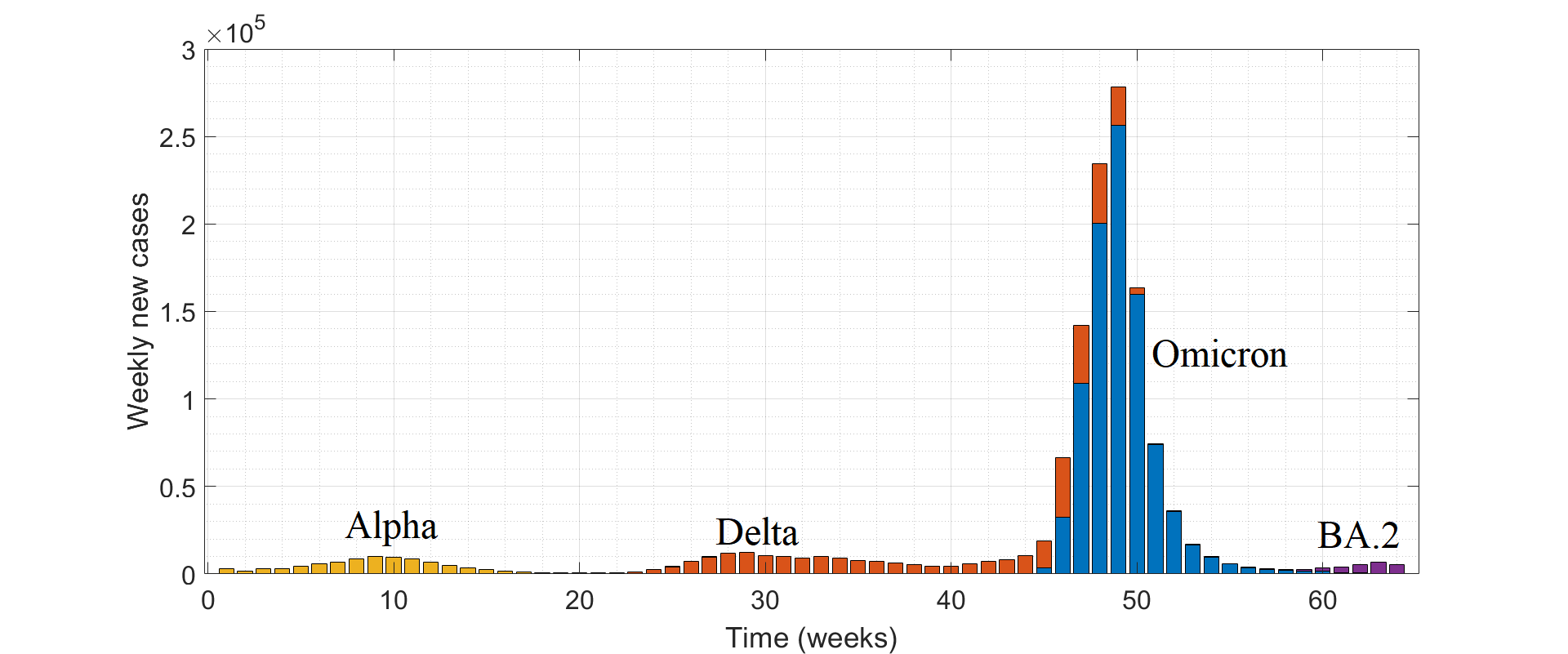}
 \caption{}
    \label{fig:New-York-weekly-variants}
\end{subfigure}
\begin{subfigure}[c]{\columnwidth}
    \centering
    \includegraphics[width=\linewidth]{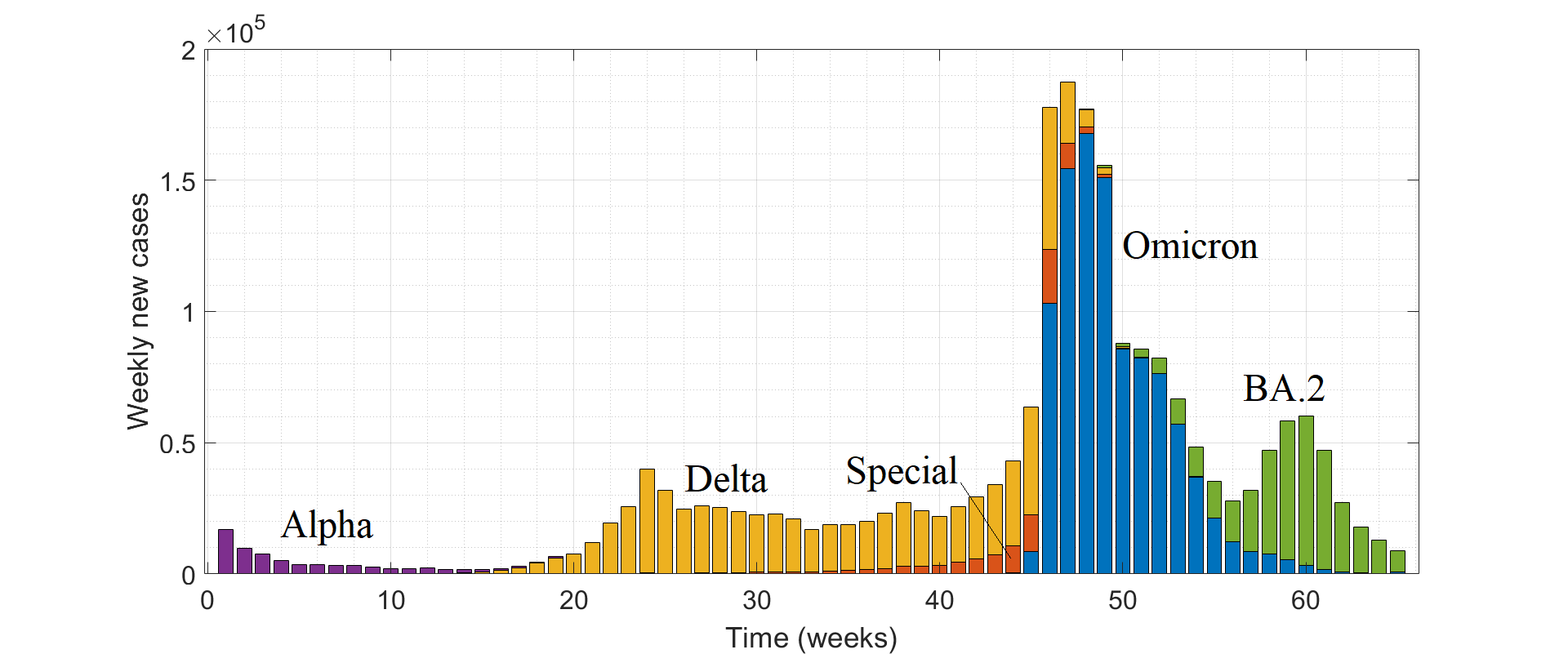}
    \caption{}
    \label{fig:London covid all}
    \end{subfigure}
    \caption{{\bf Multi-strain COVID-19 weekly case totals for NYC and London}. The data for New York City is shown in panel (a), and the data for London is shown in panel (b). The counts started when variant data was first available, January 24, 2021, for NYC and January 31, 2021, for London. The last day is April 24, 2022, when the original B1.1.529 variant was almost completely extinguished. ``Special'' denotes the Variant Under Investigation (VUI) 21-OCT-01, which spread at a rapid rate in UK but was not seen in most other countries.}
\end{figure}

Our choice of the Omicron outbreak is motivated by the fact that variant-specific testing programs were well underway when this particular variant arrived. Also, B1.1.529 was easily distinguished from the contemporaneous B1.617.2 {\blue ``Delta''} and other variants using the S-gene target failure test~\cite{mcmillen2022spike}. 
The existence of B1.1.529 and the method to detect it were 
well known well before the outbreaks in London and NYC. 
Hence, some early cases of B1.1.529 could be detected soon after the arrival of the strain in London and NYC. Furthermore, those Omicron outbreaks dwarfed those of earlier COVID-19 strains and were relatively brief, as shown in Figures~\ref{fig:New-York-weekly-variants} and \ref{fig:London covid all}. 
The weekly rate of new cases of different variants throughout 16 months of the COVID-19 pandemic for New York City is shown in Figure~\ref{fig:New-York-weekly-variants}, and for London, it is shown in Figure~\ref{fig:London covid all}. We can see a massive peak corresponding to the B1.1.529 ``Omicron'' variant around November 2021 to January 2022.

This paper presents two related approaches to fitting SIR solutions to target data. 
Our first approach is rigorous. For each $R0$, we simply identify an SIR solution using the same three major characteristics as the target outbreak: the same standard deviation, mean time, and total number of cases. We call such a pairing a ``good fit''. 
In Theorem~\ref{thm1}, we prove that 
 due to the mathematical structure of the SIR model, there exists a good fit for each value assigned to the basic reproduction number. In Theorem~\ref{thmN},
 we establish the same idea for an SIR model with $N$ compartments.

Our second approach uses a least-squares fit, numerically producing higher-quality matches. We can find a least-squares fit for each parameter value of the model. In this approach, we understand what happens with good fits, but we must resort to numerics to obtain the least-squares fits.  \black
We can also find a numerical global (absolute) least-squares fit for each parameter value $R_0$ or $\tau$. 

The goal of this paper is to present a mathematical structure, described in section~\ref{sec:model}, in which three of the four parameters correspond to the linear transformations of the ``incidence curve'' $-dS/dt$, i.e., the rate of new cases.
The remaining sections show the consequences of this structure. This structure exists for SIR partly because of our perhaps unusual form of the SIR equations, where we use the basic reproduction number as a parameter instead of the usual rate of infectious contacts per unit time. The equations are equivalent, but the scaling is different.

We report what happens when we take the straightforward path of fitting the observed data with an SIR curve using least squares. 
We found that we get high-quality fits over a range of values of the basic reproduction number and the infectiousness period that includes what most people would say are unrealistic values, and we would agree. There is reasonable agreement over a perhaps unreasonably large range of parameters.

 When we compute the best least-squares fit for each value of the average infectious period in a wide range, we obtain a relative error that is almost constant. The best least square fit suggests a mean infectiousness period is far larger than we would have expected. But we draw no biological or medical lessons from our fits about the true value of the infectious period. Rather, we want the reader to understand the mathematical structure of the SIR model and why we obtain such a large range of good fits.

\section{Model and parameters}\label{sec:model}
When the infected individuals are uniformly distributed in a population and the recovery and transmission rates are fixed, the outbreak (or epidemic) can be modeled using a standard SIR model. 
\begin{subequations}\label{eq:SIR=old}
 \begin{align}
    \dfrac{dS}{dt} = &-\beta S I
    \\
    \dfrac{dI}{dt} = & \; \beta S I
    -\dfrac{I}{\tau}   \\
     \dfrac{dR}{dt} = &\; \dfrac{I}{\tau}
\end{align}
\end{subequations}
The variables $S$ and $I$ denote the fraction of the population that is susceptible and infected, respectively.
 $R$ is the fraction removed due to recovery or death.
It is redundant since $R = 1-S-I$ and makes the system look unnecessarily intimidating.  We omit it below. The average infectiousness  {\it  per unit of time} per person is $\beta$,  and
the average duration of infectiousness is denoted by $\tau >0$. 

Model~\eqref{eq:SIR=old} is described in many studies, including \cite{miller_note_2012} and \cite{turkyilmazoglu_explicit_2021}. 
 In this paper we 
substitute the symbol $\rho$ for $\beta\tau$, so $\rho$ is the basic reproduction number $R0$.
The equations, which we will refer to as  Model~\eqref{eq:SIR},  are as follows:

\black
\begin{subequations}\label{eq:SIR}
 \begin{align}
    \dfrac{dS}{dt} = &-\dfrac{\rho S I}{\tau}\label{eq:Sa}
    \\
    \dfrac{dI}{dt} = & \; \dfrac{\rho S I-I}{\tau}   \text{ where}\label{eq:Sb} \\
    S(-\infty) = & \;1. \label{eq:S initial}
\end{align}
\end{subequations}
{Equation~\eqref{eq:S initial} implies that $I(-\infty)=0$.} 
Our form has advantages: we will see below how the solutions scale with $\tau$, and the final size of the outbreak is purely a function of $\rho$ and is independent of $\tau$ (see Equation~\eqref{rho1}). We will always assume $\rho>1$ because then and only then, Model~\eqref{eq:SIR} can sustain an outbreak.

We often denote time $t$ by a subscript, writing $I(t)$ as $I_t$, $S(t)$ as $S_t$, $S({-\infty})$ as $S_{-\infty}$, and $S({\infty})$ as $S_{\infty}$.
The condition stated as Equation~\eqref{eq:S initial} describes that the susceptible fraction at the beginning of the outbreak is 1.  In the biological literature, it is common to instead choose a time $t_0$ early in the outbreak, perhaps when the investigators first detect a case, and assume $S_{t_0} = 1$ or possibly $S_{t_0} +I_{t_0} = 1$. 
One problem with these choices is that the fraction of susceptibles who get infected before time $t_0$ and the fraction of susceptibles who get infected after time $t_0$ and up to time $t=\infty$ depend on the choice of $t_0$. 
Our choice of the initial condition $S_{-\infty} = 1$ makes some results independent of when the observers start observing and of their choice of $t_0$. For example, later in this paper, Equation~\eqref{rho1} shows that 
the final size of the outbreak is only a function of the basic reproduction number,
a result that does not depend on the choice of $t_0.$ 
Furthermore, if $S(t)$ is a solution, the solution shifted by time $\lambda$, $S(t -\lambda )$, satisfies our boundary condition but not a boundary condition such as $S(t_0)=0.9$.

Furthermore, the condition stated in Equation~\eqref{eq:S initial} implies a simpler relationship between $I_0$ and $S_0$. Noting that $S(t)$ is a monotonically decreasing function and dividing $I'$ by $S'$, we obtain  
\begin{equation}\label{dI/dS}
    \frac{dI}{dS}=-1+\frac{1}{\rho S}. 
\end{equation}
 Integrating the above equation with respect to $dS$ from $S_{t_0}$ to $S_{t}$ gives the following well-known equation, see also \cite{turkyilmazoglu_explicit_2021};
\begin{equation}\label{I=f(S)+C}
   I_{t}-I_{t_0} = S_{t_0} -S_{t} + \dfrac{1}{\rho}\ln(\dfrac{S_{t}}{S_{t_0}}).
\end{equation}

Taking the limit of the above equation~\eqref{I=f(S)+C} as $t\to-\infty$ yields
\begin{align*}
   I_{-\infty}-I_{t_0} =\; & S_{t_0} -S_{-\infty} + \dfrac{1}{\rho}\ln(\frac{S_{-\infty}}{S_{t_0}}).
\end{align*}
Since $S_{-\infty}=1$
by Equation~\eqref{eq:S initial} and  $I_{-\infty}=0$, the above equation simplifies to      
\begin{align}
 I_{t_0} =\; & 1 -S_{t_0} + 
      \dfrac{1}{\rho}\ln(S_{t_0}).\label{eq:-infinity}   
\end{align}
This equation shows for any time $t_{0}$, how $I_{t_0}$ can be determined from $S_{t_0}$ and $\rho$. 
If for a given $\rho$, scientists prefer to start a solution by specifying $S_{t_0}$, they can evaluate Equation~\eqref{eq:-infinity}  to determine the corresponding value of $I_{t_0}$.

\begin{figure}[H]
\begin{subfigure}[c]{0.5\columnwidth}
    \centering
    \includegraphics[width=0.98\columnwidth]{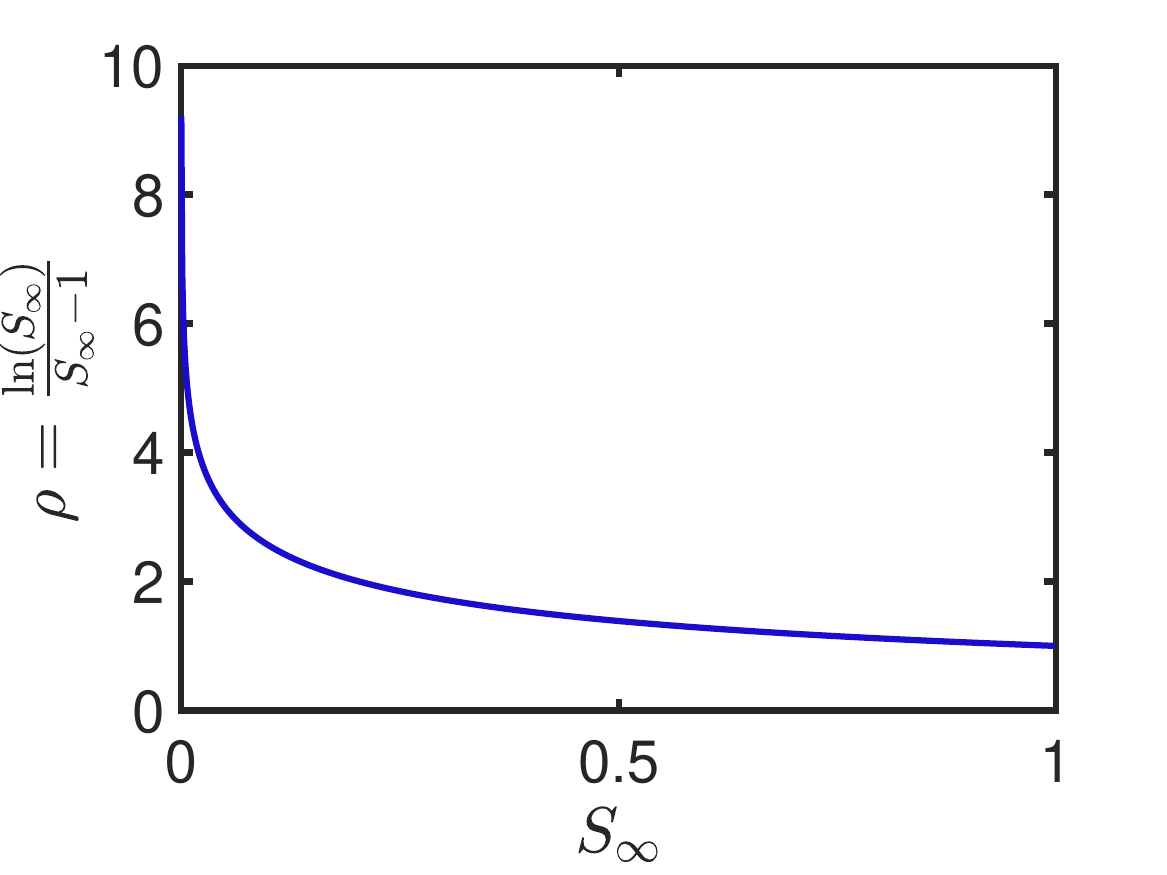}
    \caption{}
    \end{subfigure}
\begin{subfigure}[c]{0.5\columnwidth}
    \centering
    \includegraphics[width=0.98\columnwidth]{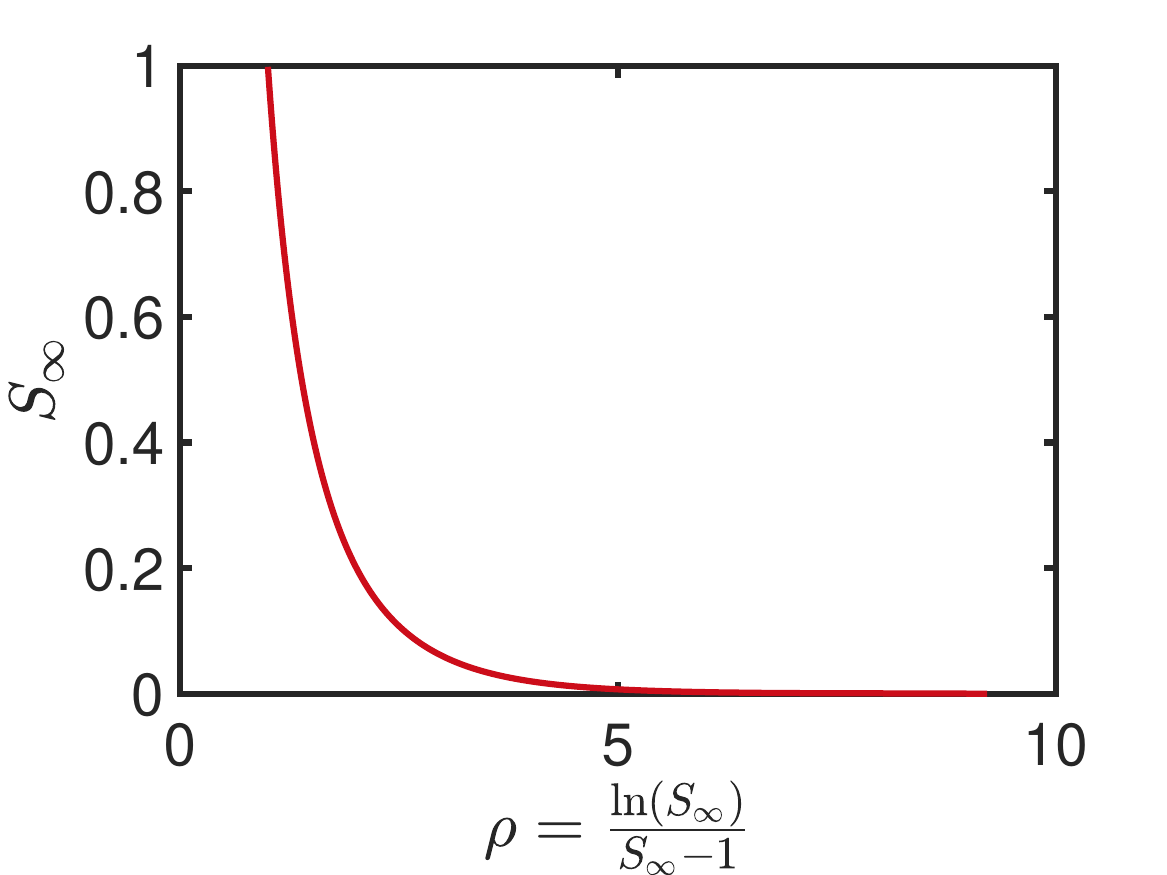}
    \caption{}
    \end{subfigure}
    \caption{{\bf  The ``final'' fraction of the initial susceptibles for an SIR outbreak who are eventually infected is $1-S_{-\infty}$.} Equation~\eqref{rho1} says that knowing $S_{-\infty}$ depends only on $\rho$, the basic reproduction number. 
    Panel(a) presents the basic reproduction number $\rho$ as a function of $S_\infty$. Panel(b) illustrates how 
    $S_\infty$ is equivalent to knowing the basic reproduction number $\rho$. 
    Since (a) shows the inverse of (b), they are the same curve with the axes switched.}
    \label{fig:S at infinity-Vs-Rho}
\end{figure}

An SIR model's outbreak has a ``{\bf final size}'' denoted as $\Delta = S_{-\infty}-S_{\infty}$. By Equation~\eqref{eq:S initial}, $\Delta = 1-S_{\infty}$. 
By setting $S_{t_0}=S_\infty$ (\ie,  $t_0=\infty$) and $I_\infty=0$
in Equation~\eqref{eq:-infinity}, we obtain a monotonic relationship between $\rho$ and $S_{\infty}$,
\begin{align} \label{rho1}
 \rho=& \dfrac{\ln(S_{\infty})}{S_{\infty}-1}.
 \end{align}
The above equation is discussed in the literature~\cite{miller_note_2012,brauer2019early}. Equation~\eqref{rho1} implies that the final size of an outbreak only depends on the basic reproduction number $\rho$; see Figure~\ref{fig:S at infinity-Vs-Rho} for an illustration. 
Note that Equation~\eqref{rho1} is defined for $S_\infty \in (0,1)$ as shown in Figure~\ref{fig:S at infinity-Vs-Rho}. The final size, $\Delta(\rho)$, is monotonically increasing as $\rho$ increases and 
$\Delta(\rho)\to 1$ as $\rho\to \infty$.
 See Figure~\ref{fig:S at infinity-Vs-Rho}(b).

\begin{figure}[H]
\begin{subfigure}[c]{0.5\columnwidth}
        \centering
    \includegraphics[width=0.9\columnwidth]{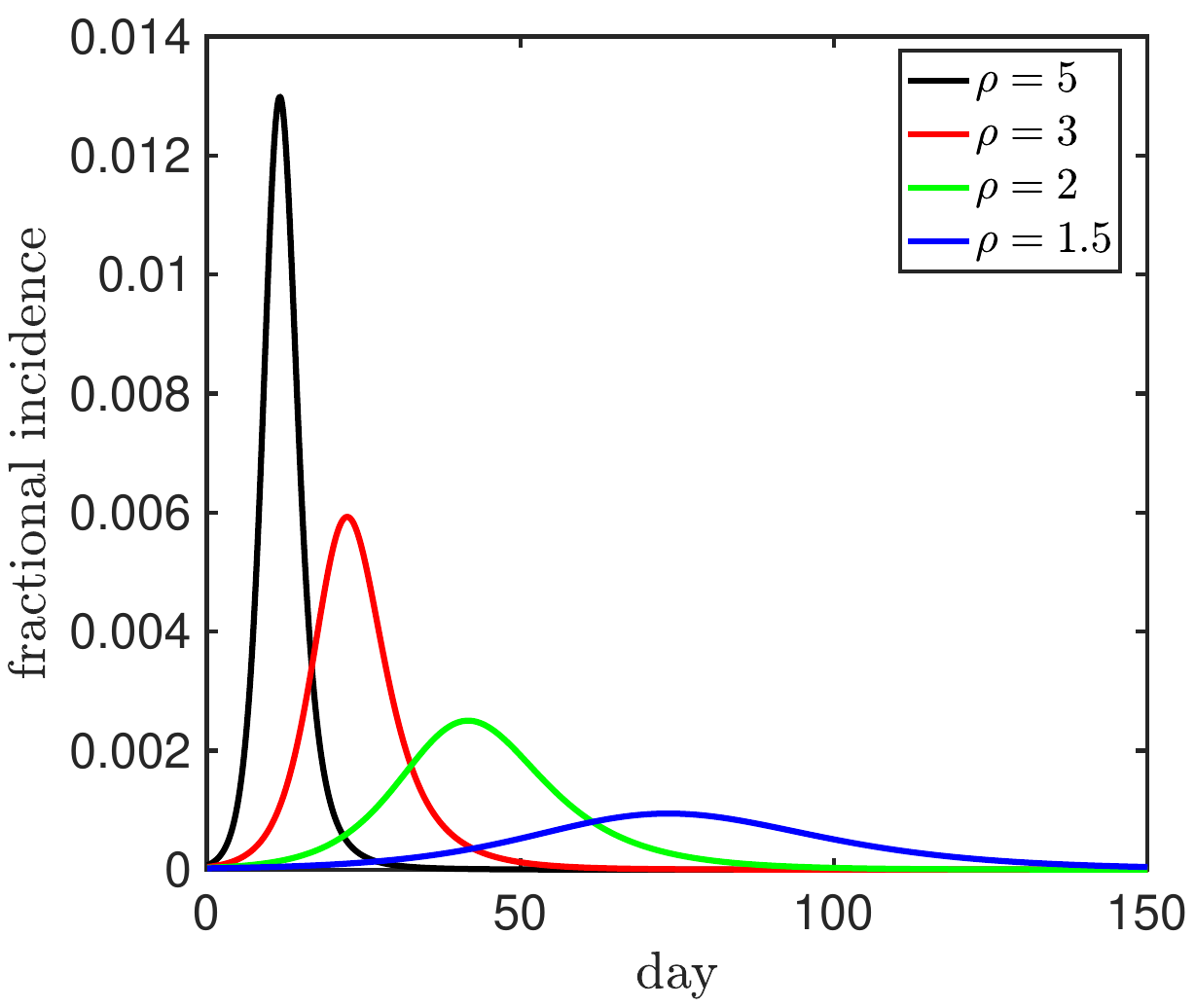}
    \caption{\red }
\end{subfigure}
\begin{subfigure}[c]{0.5\columnwidth}
        \centering
    \includegraphics[width=0.9\columnwidth]{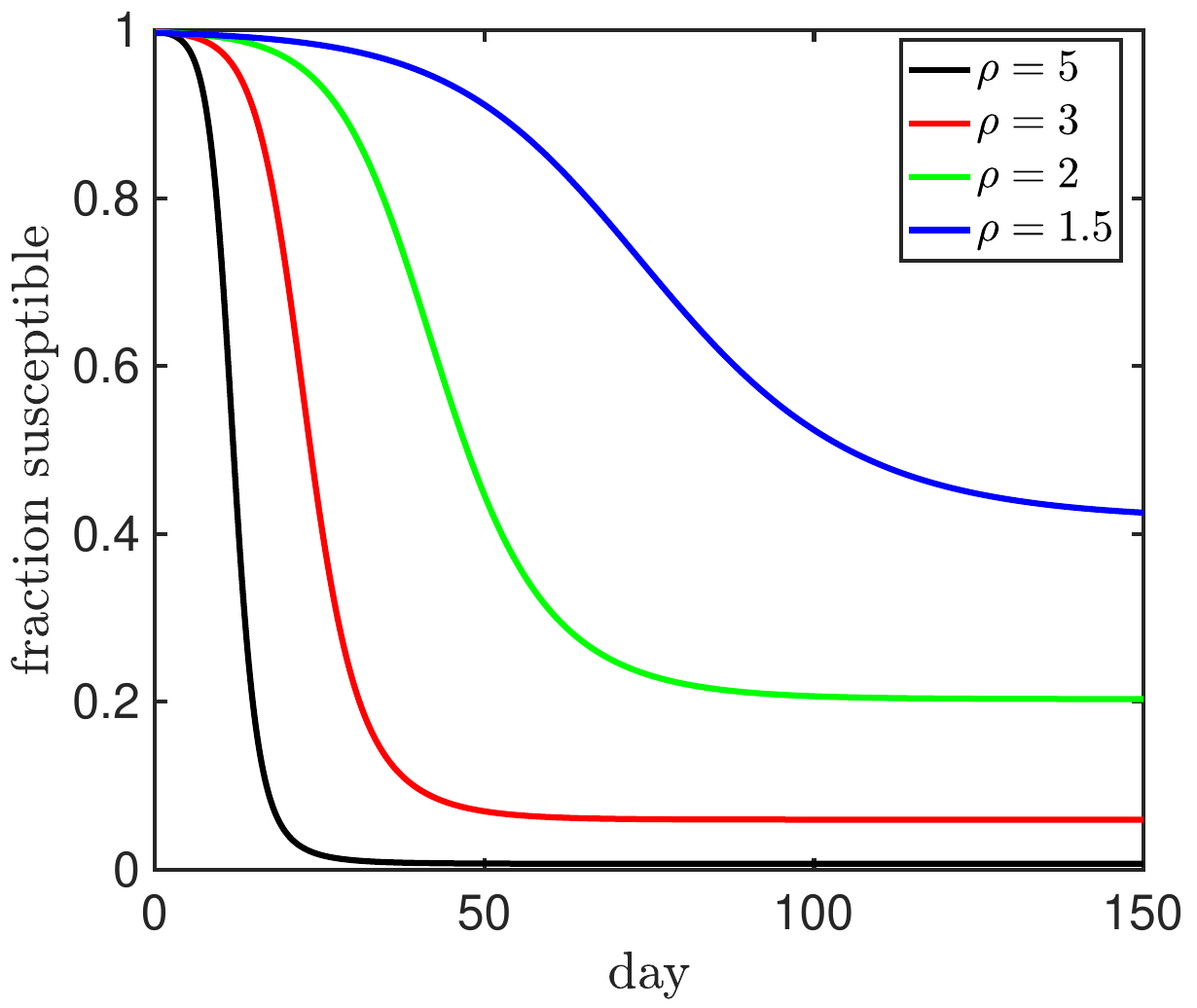}
    \caption{}
\end{subfigure}
    \caption{{\bf\BF SIR Incidence $-S'(t)$  and susceptible fractions $S(t)$ for four examples.} The four example outbreaks
for Model \eqref{eq:SIR} have the same $\tau=7$ and different $\rho$ values as shown in the panels. They are initialized by setting $1-S(0)= 0.001$, and I(0) is calculated from Equation~\eqref{eq:-infinity}, which implies that $S(-\infty)=1$.
    Panel (a) shows the fractional incidence $-S'(t)$ for solutions.
    Panel (b) shows the susceptible fraction
    for the corresponding outbreaks 
    in Panel (a).}
    \label{fig:SIR different rhos}
\end{figure}

{\bf\BF Using $-S'$ to fit outbreak data requires four parameters.}
The fractional incidence (the fractional rate of new cases per unit time) is $-S'$ where $S' = dS/dt$, shown in Figure~\ref{fig:SIR different rhos}.
 In addition to $\rho$ and $\tau$, there are two additional parameters, the mean time of the outbreak $\mu$, and adjustment factor $A$ for the size of the outbreak, both of which we now define. 
 
 The mean time $\mu_f$ of an integrable function $f$ is $\mu_f=\frac{\int_\R tf(t)dt}{\int_\R f(t)dt}$. 
 For the mean time $\mu$ of the outbreak's incidence, let $f=-S'$. Hence 
 $$ \mu=\frac{ \int tS'(t)dt}{\int S'(t) dt}. $$ 
 
 The multiplicative parameter $A$ adjusts 
the size of an outbreak, \ie, the total number of people infected, according to the outbreak data, and $-AS'$ represents the rate of new cases per day. 

{\bf\BF Using the multiplicative factor $A$ to adjust the size of the outbreak.}
Let $\Delta$ denote the (fractional) final size of an outbreak, $\int_\R -S'=1-S_\infty$, which is less than one. For cities like NYC or London, the reported final size may be approximately $10^6$. To adapt the final size to that scenario, the value of $A$ would be set to $10^6/\Delta$. 
In this way, the term $\int_\R -A S'$ would yield a final size of $10^6$ for the model. (In general, $A$ is the ratio of the final sizes of two outbreaks.)
Alternatively, to compare two SIR solutions that have different $\rho$ values and final size values $\Delta_1$ and $\Delta_2$, we can adjust the final size of the second model by scaling its $-S'$ term to  $-AS'$ using a factor $A=\Delta_1/\Delta_2$. This adjustment ensures that the second model has a final size of $\Delta_1$. \black

A difficulty with fitting can be seen when we fit the Omicron outbreak in NYC; there are approximately 1 million total case reports in a population of about 8 million (see Supplementary Material, Sec.~\ref{supappendix data}). 
Total case reports are an unknown fraction of all cases.
Hence, we do not choose a best estimate for $\rho$ for this outbreak. In addition, we do not know how much immunity there was in the population due to immunizations and prior COVID-19 infections.

\section{Our SIR principle of three linear (or affine) scalings}\label{sec good fits}
In this section, we focus on our first approach of fitting an SIR model incidence curve $-S'(t)$ to incidence data. In order to do that, we define a few terms that we use in this paper. 

{\begin{Def}[Target triple] \rm
 Let $f:\R\to[0,\infty)$  be an integrable function. Assume $A_f:=\int_\R f(t)dt \in (0,\infty)$, $\mu_f:= 
  \frac{\int_\R tf(t)dt}{\int_\R f(t)dt}
=
 \frac{\int_\R tf(t)dt}{A_f} \in \R$, and $\sigma_f:= \Big(\frac{\int_\R (t-\mu_f)^2f(t)dt}{A_f}\Big)^{1/2} \in (0,\infty)$. 
 In other words, let $A_{f}$, $\mu_{f}$, and $\sigma_{f}$ denote the total outbreak size, the mean time, and the standard deviation of a given function $f$, (where usually $f=-S'$). We call such a triple
$(A_f,\mu_f,\sigma_f)$ a {\bf target triple}. 
\end{Def}}

{\begin{Def}[good fit]\rm
For a target triple $(A_f,\mu_f,\sigma_f)$, we
say a function $g:\R\to[0,\infty)$ 
is a {\bf good fit} (for the target triple) if $(A_g,\mu_g,\sigma_g)=
(A_f,\mu_f,\sigma_f)$.
\end{Def}}
Sometimes we write Model~(\ref{eq:SIR};$\rho,\tau$) to indicate which values are assigned to parameters of Model~\eqref{eq:SIR}. Theorem~\ref{thm1} says that for each $\rho \in (0, \infty)$ there is a good fit. A bit of caution is appropriate: ``Good fit" does not mean ``great fit".  
 \begin{thm}\label{thm1}{\rm
  Assume there exists a non-constant solution  $(S(t),I(t))$  of Model~(\ref{eq:SIR};$\rho,\tau=1$)  for which $-S'$ has mean time $\mu=0$. Let
    $(A_f,\mu_f,\sigma_f)$ be a target triple. Let $\rho >1$. There exist $A>0$, $\tau >0$ and a 
    solution $(S_1,I_1)$ of Model~(\ref{eq:SIR};$\rho,\tau$) such that the resulting $-S_1'$ multiplied by $A$ is a good fit to $(A_f,\mu_f,\sigma_f)$. 
}\end{thm}

First we establish some useful facts.
{\begin{proposition}\label{prop1} \rm
Assume there exists a non-constant solution  $(S(t),I(t))$  of Model~(\ref{eq:SIR};$\rho,\tau=1$)  for which $-S'$ has mean time $\mu=0$. The following are true:
\begin{enumerate}[label=(P\arabic*), ref=(P\arabic*), wide, labelwidth=0pt, labelindent=0pt]
\item
For every $\aaa>0$ and $\mu$,  
$(S(\frac{t-\mu}{\aaa}),I(\frac{t-\mu}{\aaa}))$ 
is a solution of Model~(\ref{eq:SIR};$\rho,\tau=\aaa$). \label{Propo:1}\\
\item Let $f(t)=-S'(\frac{t-\mu}{\aaa})$. Its mean time is $\mu$, and its standard deviation is $\sigma\cdot\aaa$, where $\sigma$ is the standard deviation of $-S'(t)$.\label{Propo:2}
\end{enumerate}

\end{proposition}}

\begin{proof}
To prove statement \ref{Propo:1}, note that since $(S(t),I(t))$ is a solution of Model~(\ref{eq:SIR};$\rho,\tau=1$), so by Equation~\eqref{eq:Sa}
 $$
S'(t)=\frac{d}{dt}S(t) = -\rho S(t) I(t).
$$
 Therefore 
 $(S(\frac{t-\mu}{\aaa}),I(\frac{t-\mu}{\aaa}))$ 
 satisfies 
{$$
\frac{d}{dt}S\Big(\frac{t-\mu}{\aaa}\Big) = 
\frac{S'(\frac{t-\mu}{\aaa})}{\aaa} =\frac{
-\rho S(\frac{t-\mu}{\aaa}) I(\frac{t-\mu}{\aaa})
}{\aaa},
$$}
so $S(\frac{t-\mu}{\aaa})$ satisfies Equation~\eqref{eq:Sa} with $\tau =\aaa$. Similarly $I(\frac{t-\mu}{\aaa})$ satisfies Equation~\eqref{eq:Sb} when $\tau=\aaa>0$. Furthermore
Equation~\eqref{eq:S initial} is satisfied, proving that $(S(\frac{t-\mu}{\aaa}),I(\frac{t-\mu}{\aaa}))$ is a solution of Model~(\ref{eq:SIR};$\rho,\tau= a$).

To prove statement \ref{Propo:2} about the mean,
apply a change of variables $x=\frac{t-\mu}{\tau}$ or $t = x\tau+\mu$. 
\begin{align}
    \mu_f= &
  \frac{\int_\R tf(t)dt}{\int_\R f(t)dt}
=   \frac{\int_\R tS'(\frac{t-\mu}{\tau})dt}{\int_\R S'(\frac{t-\mu}{\tau})dt}\\
= &  \frac{\int_\R (x\tau+\mu) S'(x)\tau dx}{\int_\R S'(x)\tau dx}
=   \tau\frac{ \int_\R xS'(x) dx}{\int_\R S'(x) dx} + \mu =\tau\cdot0 +\mu = 0+\mu\\
\end{align}
The claim about the standard deviation in \ref{Propo:2} follows from a similar calculation and is omitted here.
 \end{proof}
Note that if $(S(t),I(t))$ is {\em any} non-constant solution of Model~(\ref{eq:SIR};$\rho,\tau=1$) with the mean time $\mu \neq 0$,
then $(S(t+\mu),I(t+\mu))$ is also a solution of Model~(\ref{eq:SIR}) with the same $\rho$ and $\tau=1$, and the mean time of the outbreak  $-S'(t+\mu)$ is $0$. 
Hence, in the assumption of the above proposition, the existence of a solution with a finite mean time is required and the value of the mean time is not important.
The fact that $-S'$ has a finite mean time follows from the fact that $-S'(t) \to 0$ exponentially fast as $|t| \to \infty$.
 
\begin{proof}[Proof of Theorem~\ref{thm1}.]
Let $(A_f,\mu_f,\sigma_f)$ be a target triple. By Proposition~\ref{prop1}, $(S(\frac{t-\mu_f}{\tau}),I(\frac{t-\mu_f}{\tau}))$ is a solution of Model~(\ref{eq:SIR};$\rho,\tau$) that $-S'(\frac{t-\mu_f}{\tau})$ has mean time $\mu_f$ and standard deviation $\sigma_1 \cdot \tau$, where $\sigma_1$ is the standard deviation of $-S'(t)$.

Let  $\tau= \frac{\sigma_f}{\sigma_1}$. Let $(S_1(t), I_1(t))=
(S(\frac{t -\mu_f}{\tau}),I(\frac{t-\mu_f}{\tau}))$.
Then $(S_1(t), I_1(t))$ is a solution of 
Model~(\ref{eq:SIR};$\rho,\tau$), and its incidence 
has mean time $\mu_f$ and standard deviation $\sigma_f$.

Let $A = \frac{A_f}{-\int_\R S'_1(t)dt}$. Then 
 $-AS_1'(t)$ has integral equal to $A_f$ and $-AS_1'$ is a good fit for the 
target triple.
\end{proof}


{\bf Finding even more good fits when using multicompartmental SIR models.} 
When we explained the above findings to our colleagues, we repeatedly heard the suggestion that perhaps the problem was that the SIR model is too simple. It does not include enough detail, that perhaps we should add parameters or variables to allow for subpopulations with different susceptibilities, or perhaps different levels of infectiousness, or different sociablility. Such details do exist in actual populations. So here we report on what we found.

More complex models allow splitting the population into subgroups, each having its own susceptible and infected fractions and infectious period. 
For example, we might split the population into five age categories, and we might assume each category has two intervals of susceptibility since some people might have previous exposures, and we might split the population into three categories according to how many social contacts they have. That means that in this somewhat arbitrary example, there are 
$N=5 \times 2 \times3$ compartments.

In general, let $N$ be the number of subpopulations (or compartments). The susceptibles and infectious are $N$-vectors whose components are fractions of the total population, 
$\vec{S}=(S_1,\ldots,S_N)$ and $\vec{I}=(I_1,\ldots,I_N)$.
We will let $i,j$ be coordinates, $1\le i,j\le N$.
Each $S_i$ (or $I_i$) is the fraction of group $i$ that is susceptible (infectious, respectively) and  $S_i(-\infty)= 1$. The mean duration of infectiousness of $I_i$ is denoted by $\tau_i$. 
There are $N^2$ basic reproduction numbers
$\rho_{i,j}>0$, the number of people in group $i$ who will be potentially infected by an infected person in group $j$, and $S_i  \frac{\rho_{i,j}}{\tau_j}I_j$ is the fractional rate of new cases in group $i$ caused by infectious people in group $j$.  We do not explicitly specify the relative sizes of the groups.
Each number $\rho_{i,j}$ depends in part on the sizes of groups $i$ and $j$, and because they are transmission rates, 
they are extremely hard to estimate. Errors in these numbers might be systematically high (or low) and the errors would compound and not cancel each other. The SIR model that describes infection transmission is as follows 
\begin{subequations}\label{eq:SIR-N}
\begin{align}
    \frac{d }{dt}S_i = &-S_i \sum_{j=1}^N \frac{\rho_{i,j}}{\tau_{j}}I_j 
   \\
    \frac{d}{dt}I_i = &\ S_i \sum_{j=1}^N \frac{\rho_{i,j}}{\tau_{j}}I_j-\frac{I_i}{\tau_{i}}       \text{ where} \\
    S_i(-\infty) = & 1, i= 1,\ldots,N. \label{S-N initial}
    \end{align}
\end{subequations}

Theorem~\ref{thmN} below is an analog of Theorem~\ref{thm1}, with many free transmission parameters, analogues of $\rho$ and $\tau$. 

While $\rho>1$ is needed in Theorem~\ref{thm1}, for Model~\eqref{eq:SIR} to have an outbreak, we do not have lower bounds for the $N^2$ values $\rho_{i,j}$, since an outbreak can also depend on the relative sizes of $\tau_i$. 
It is sufficient to have 
$\rho_{i,i}>1$ for some $i$ to have an outbreak, since then the $i^{th}$ group would support an outbreak by itself.

Let $p_i > 0$ be the fraction of the total population in group $i$, so $\sum p_i =1$.
When using a solution $(\vec{S},\vec{I})$ to fit an outbreak, we use the fractional rate of new cases, namely 
$-\sum_i p_i S'_i$. 
For $A>0$,  we say $g(t):=-A\sum_i p_i S'_i$ is a {\bf good fit} to
$(A_f,\mu_f,\sigma_f)$
if $\int_{-\infty}^{\infty} g = A_f$, the mean and the standard deviation of $g$ are $\mu_f$ and $\sigma_f$.

For cases where $N=30$ as described above, there is a 900-dimensional set of matrices $M$ that yield good fits to each target. When
$M$ does not have a real eigenvalue $>0$,
there is no outbreak, so there is no good fit, since then each solution of 
Model~\eqref{eq:SIR-N} has $I(t)=0$ for all $t$.

For simplicity to specify 
which values of $(\rho_{i,j})$ and $\tau_i$ a solution of Model~\eqref{eq:SIR-N} depends upon, we write $(\vec{S},\vec{I})(t)$ as $(\vec{S},\vec{I})(t;(\rho_{i,j});\tau_1,\dots,\tau_N)$, or that it is a solution
of Model~(\ref{eq:SIR-N};$(\rho_{i,j});\tau_1,\dots,\tau_N$)
\begin{thm}[An $N^2$-dimensional set of good fits to each target triple]\label{thmN} {\rm 
Let $(A_f,\mu_f,\sigma_f)$ be a target triple. 
Assume there exists  a nonconstant solution $(\vec{S},\vec{I})(t)$ of 
Model~(\ref{eq:SIR-N};$(\rho_{i,j}); \tau_1,\cdots,\tau_N$) for some $(\rho_{i,j})$ and $(\tau_1,\cdots,\tau_N)$ and that its incidence has finite mean time and finite standard deviation. There exist constants $A,\phi>0$ and  
a solution 
$(\vec{S_1},\vec{I_1})(t)$ of
Model~(\ref{eq:SIR-N}; $(\rho_{j,k}); \phi\tau_1, \cdots, \phi\tau_N$) such that $g(t):=-A\sum_i p_i S'_i$ is a good fit to $(A_f,\mu_f,\sigma_f)$.}
 \end{thm}
We conjecture that for every non-constant solution $(\vec{S},\vec{I})(t)$
the incidence $g(t)$ has finite mean time and finite standard deviation.

 \begin{proof}
Let $\mu_0$ and $\sigma_0$ denote the mean and the standard deviation of the solution $(\vec{S},\vec{I})(t)$. 
 Let $ \phi= \frac{\sigma_f}{\sigma_0}$. 
$(\vec S_\phi(t),\vec I_\phi(t))=(\vec S(\frac{t}{\phi}),\vec I(\frac{t}{\phi}))$ is a solution to Model~(\ref{eq:SIR-N}; $(\rho_{j,k}); \phi\tau_1, \cdots, \phi\tau_N$) such that $-\sum_i p_i S'_{\phi i}$ has mean time $\mu_0 \phi$ and standard deviation $\sigma_0 \phi =\sigma_f$.
   
Let $\lambda =\mu_f - \mu_0 \phi$. $(\vec S_\lambda^{*}(t),\vec I_\lambda^{*}(t))= (\vec S_\phi(t-\lambda),\vec I_\phi(t-\lambda))$ is a solution to Model~(\ref{eq:SIR-N}; $(\rho_{j,k}); \phi\tau_1, \cdots, \phi\tau_N$) such that 
$-\sum_i p_i S^{*'}_{\lambda i}$ has mean time $\mu_f$ and standard deviation $\sigma_f$. Let $A = \frac{A_f}{-\sum_i\int p_i S^{*'}_{\lambda i} dt}$. Then the adjusted incidence $-A\sum_i p_i S^{*'}_{\lambda i}$ is $A_f$ and that completes the proof. 
\end{proof}

\subsection
{ Least-squares fits: finding out how good the good fits are}
\label{subsec:least squares}

 We have defined a ``good fit'' between two functions in terms of three parameters -- the amplitude, the mean, and the standard deviation. With the fitting criterion thus defined, we have shown above that there is a one-parameter family of good fits. This family can be parameterized by either $\rho>1$ or $\tau>0$. The question is: how good are these good fits?
For that, we apply least squares to measure how good the fit is.

\begin{figure}[H]
\begin{subfigure}[c]{0.5\columnwidth}
    \centering
    \includegraphics[width=\columnwidth]{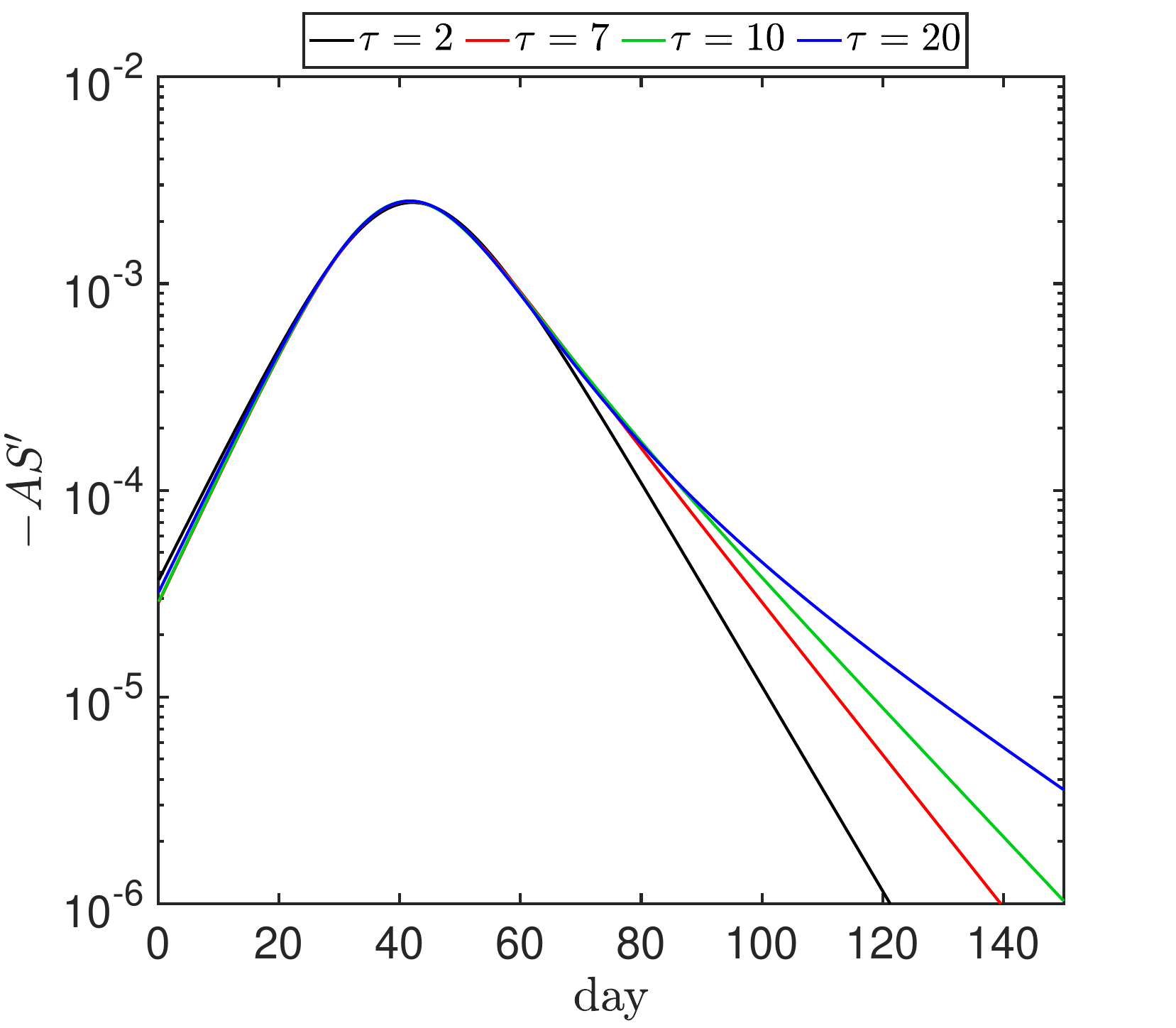}
    \caption{}
\end{subfigure}
\begin{subfigure}[c]{0.5\columnwidth}
    \centering
    \includegraphics[width=\columnwidth]{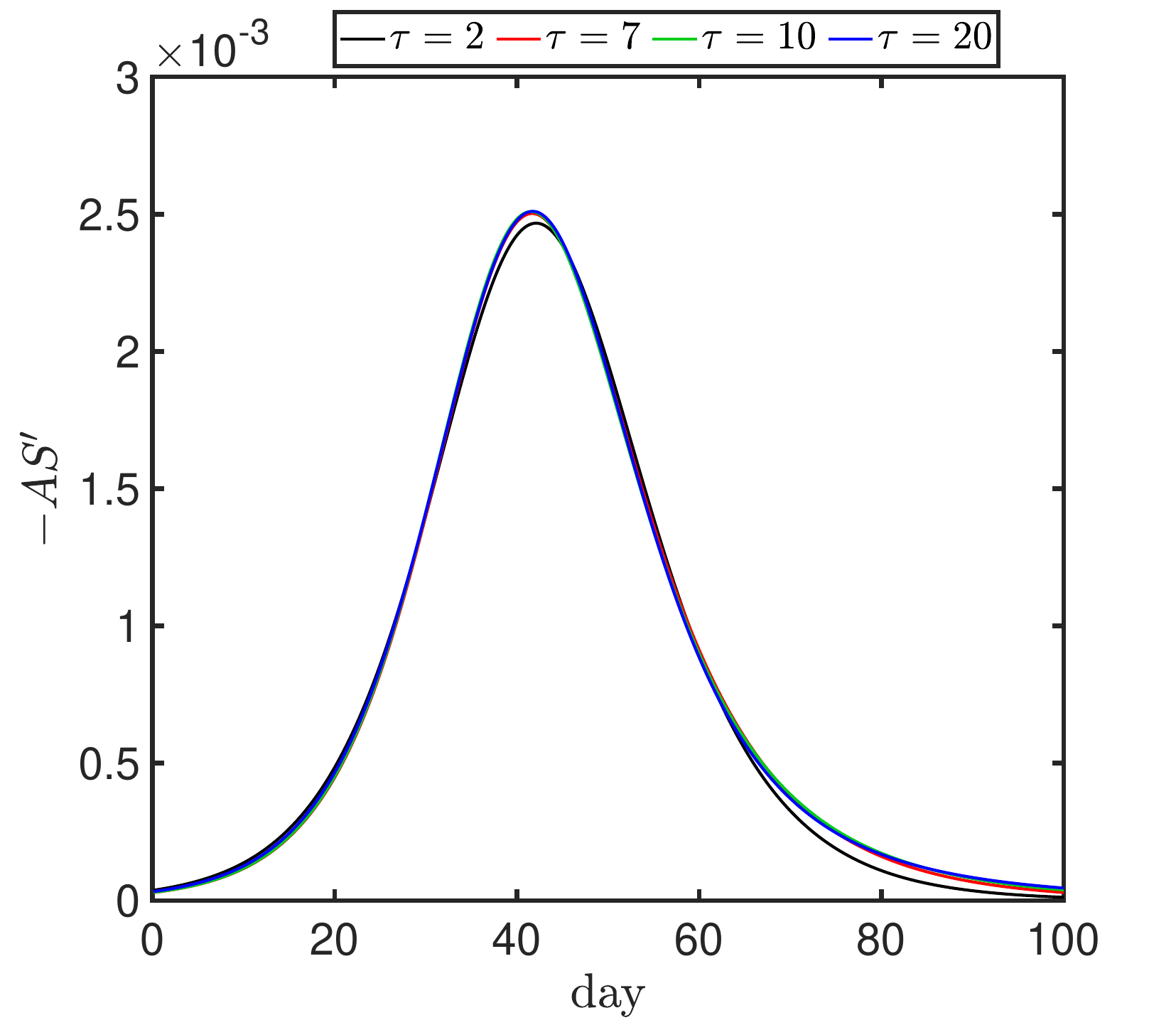}
    \caption{}
\end{subfigure}
\caption{ {\bf {\blue Least squares} fits of SIR solutions to another SIR solution}. The simulated data is obtained by solving
Model~\eqref{eq:SIR} for
$\rho = 2$ and $\tau = 7$. 
Then we find three least-squares fits to that solution using $\tau = 2,20$, and $20$.
For each optimization, we find the best values for the parameters $\rho$, $I_0$, and $A$ such that the relative error is minimized; see Table~\ref{suptable:SIR-vs-SIR} in Supplementary Material for the values of parameters in each case. Both panels show the same solutions.
The point of these plots is to show that the four solutions agree well with each other except perhaps in the region late in the outbreak where the differences can only be seen clearly in the log scale used in the left panel. }
\label{fit-four-various-tau}
\end{figure}

\begin{Def}[relative error] {\rm
   Suppose $x(n)$ and $y(n)$ are two representations of the daily new cases in day $n$ of an outbreak;
one might be actual data of new cases and the other a case history obtained from a mathematical model. We need a formula that tells how similar $y$ is to $x$. We say that the fit incidence $y$ differs from the target incidence $x$ by 
 the {\bf
 Relative Error (RE)},
\begin{equation}\label{RE}
 \err(x(\cdot),y(\cdot)) \bydef 
\dfrac{\|x-y\|}{\|x\|}
=\qty[ \dfrac{ \displaystyle{\sum_n [x(n)-y(n)]^2}}{\displaystyle{\sum_n x(n)^2}}]^{\dfrac{1}{2}},
\end{equation}
 where $\|\cdot\|$ denotes the Euclidean norm.
 For continuous time, we use the integral version of Equation~\eqref{RE}. }
\end{Def}
In this paper, the target $x(t)$ represents the daily new-case rate of an outbreak or an SIR solution.  The fit $y$ is always $-AS'(t)$ where $S(t)$ is a solution of Model \eqref{eq:SIR} for some choice of parameters. We have four parameters $\rho,\tau, I_0=I_{t_0}$, and $A$. We define $\min RE(x,y;\text{fixed})$ where ``fixed'' denotes the set of parameters if any, whose values are held fixed in $y$, and the minimum is computed by optimizing over the rest of the parameters. 
For example in Figure~\ref{fit-four-various-tau}, the red curve, which is a solution to Model~\eqref{eq:SIR} with $\tau=7$ and $\rho=2$, is chosen as the target and the black, $y(t;\tau=2)$,  the green, $y(t;\tau=10)$, 
and the blue, $y(t;\tau=20)$, curves are least squares fits when $\tau$ is held fixed and the optimization is conducted by optimizing over $\rho, I_0$, and $A$. Once the optimization function suggests a value for $I_0$, the value of $S_0$ is calculated using Equation~\eqref{eq:-infinity}.  See Table~\ref{suptable:SIR-vs-SIR} in Supplementary Material, Section~\ref{app:tables} for the parameter values.

We use the Matlab built-in function {\bf fmincon} for optimization. We use the default optimization algorithm, which uses an ``{\bf interior-point}'' approach. This algorithm can be used to solve both linear and nonlinear optimization problems. We also use constraint optimization by setting the lower and upper bounds for our parameters. The lower bounds are set to make sure the estimated parameter values are positive, as all our parameters are positive. The objective function that the optimization algorithm is minimizing is the relative error of a target data set and a solution to Model~\eqref{eq:SIR}.

\begin{figure}[H]
\begin{subfigure}[c]{0.5\columnwidth}
        \centering
    \includegraphics[width=0.8\columnwidth]{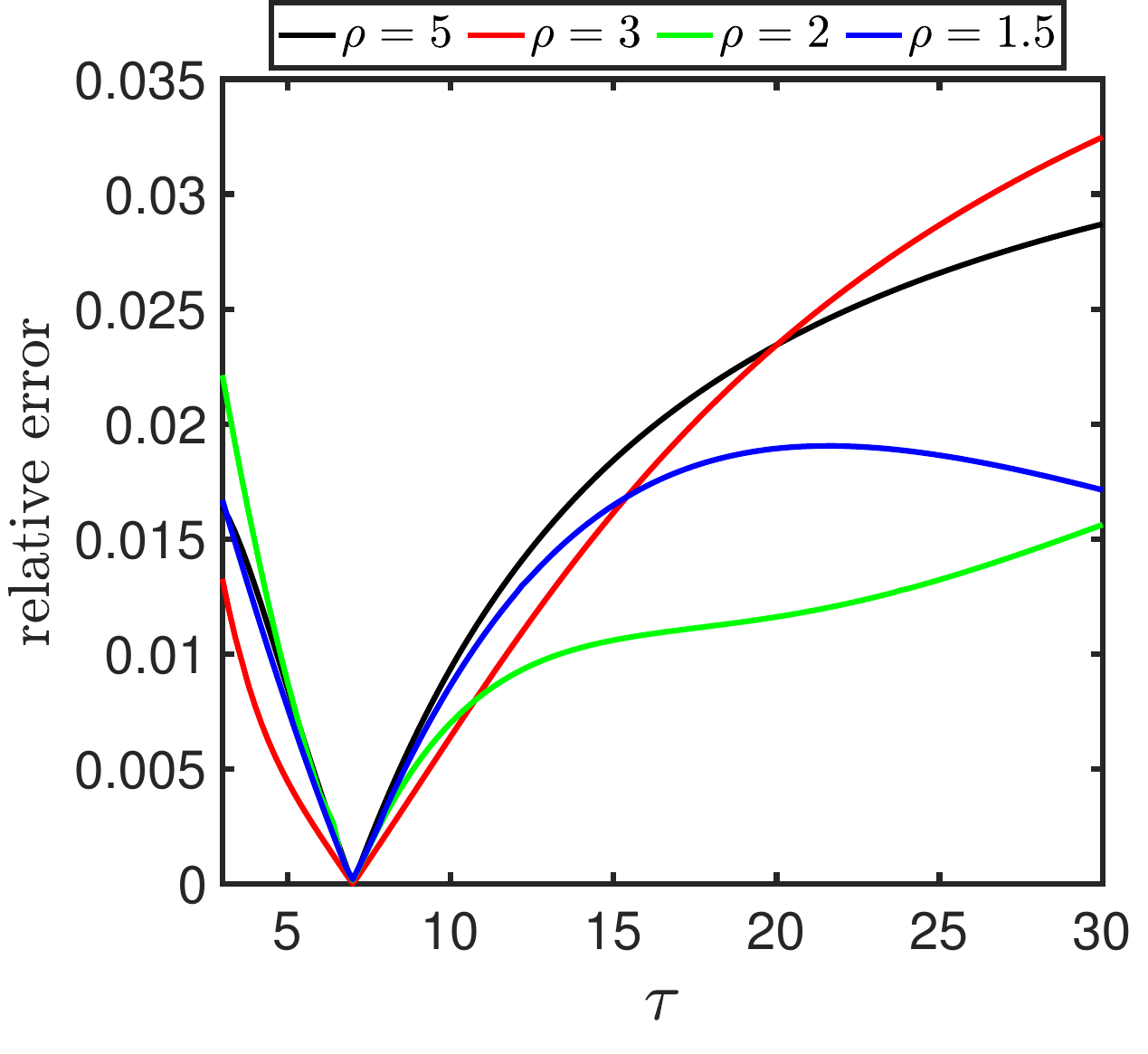}
    \caption{ }
\end{subfigure}
\begin{subfigure}[c]{0.5\columnwidth}
        \centering
    \includegraphics[width=0.85\columnwidth]{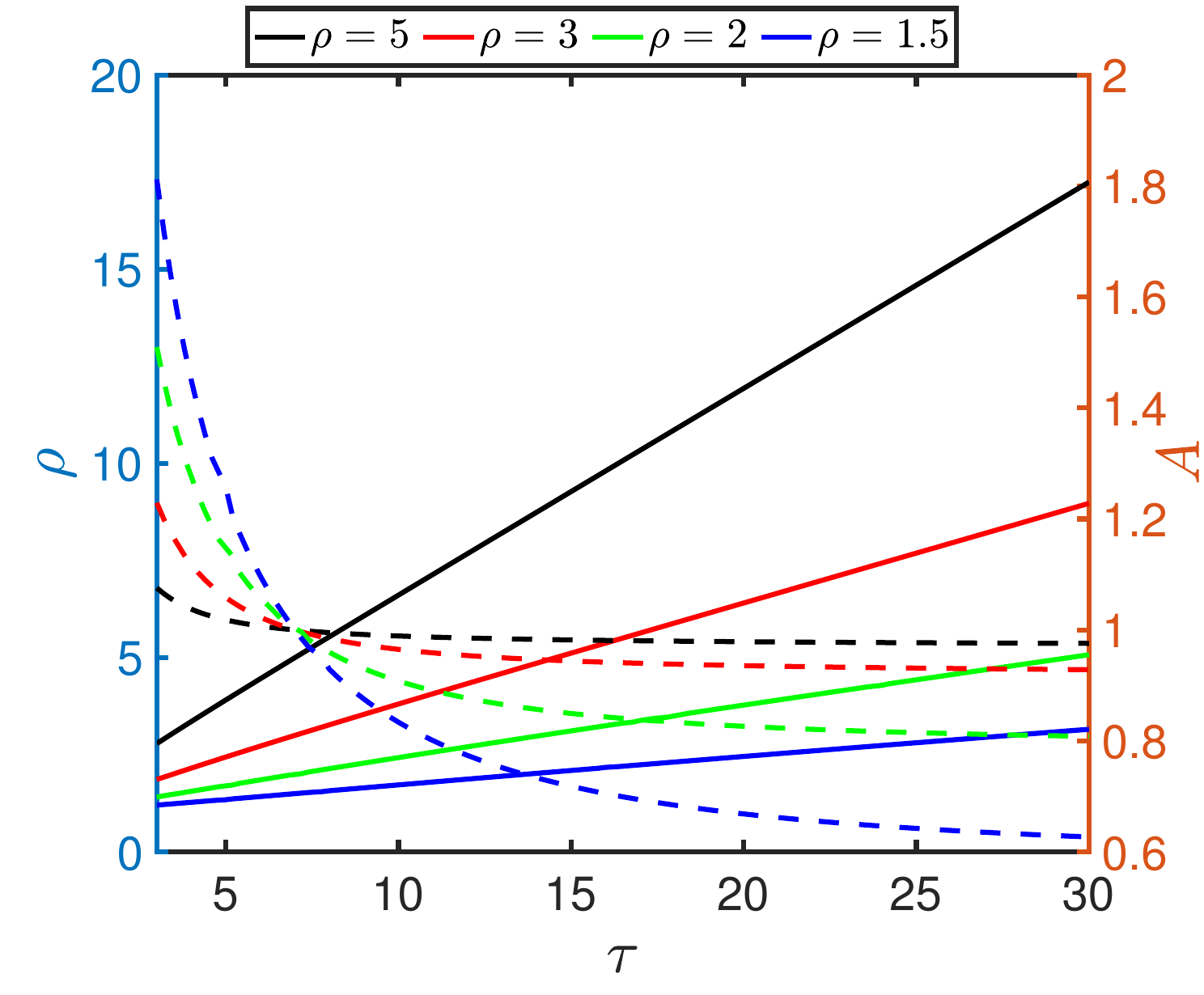}
    \caption{}
    \label{fig:comp rho tau N}
\end{subfigure}
\caption{{\bf Four families (or curves) of lease-squares fits of SIR solutions to four specific SIR simulated outbreaks.} We choose four solutions of Model~\eqref{eq:SIR} all with $\tau =7$ and the  $\rho$ values $ 1.5, 2, 3, 5$. For each, 
we use its incidence $-S'(t)$ as a target data set.
Using least squares, we fit the incidence rate of solutions of Model~\eqref{eq:SIR} to the four different target data.
For each of the four $\rho$ target data sets and $\tau$ from $3$ to $30$, we determine the least squares best set of parameters $\rho, A$, and $I_0$ for which the incidence of the Model~\eqref{eq:SIR} solution has the smallest relative error. 
{\bf Panel (a):}  The relative error is shown.
{\bf Panel (b):} $\rho$ and $A$ as a function of $\tau$ are shown in panel(b); solid lines ($\rho$) use the scale on the left, dashed lines ($A$), the scale on the right. 
Note that the global minimum of the relative error is 0 at $\tau = 7$, which is the corresponding value of $\tau$ for target data sets, as expected. The four Model~\eqref{eq:SIR} targets $-S'$ have the same $\tau=7$ and different $\rho$ values as shown in the panels. 
They are initialized by setting 
$1-S(0)= 0.001$, and 
then $I(0)$ is calculated from Equation~\eqref{eq:-infinity}, 
which implies that 
$S(-\infty)=1$.}\label{fit-four-various-rho}
\end{figure}

As discussed above in Section~\ref{sec:model}, an SIR solution 
$(S,I)$ of Model~\eqref{eq:SIR} 
is determined by four parameters. There is a quote from John von Neumann that says ``With four parameters, I can fit an elephant, and with five, I can make him wiggle his trunk" (Freeman Dyson \cite{dyson2004meeting}). The implication is that using too many parameters could be an over-parametrization where the parameter values might not correspond to the true values of a physical situation. (Note: von Neumann was speaking of complex parameters.)
In particular there might be other choices of the parameters that give a similar result.
 Such a qualitative statement suggests we should be careful whenever we wish to determine parameters from the least {\blue squares fit}, for any epidemic model. For example, in Figure \ref{fig:NYC best fit}, two SIR least squares fits to NYC Omicron outbreak data are shown. These two SIR outbreaks are much more similar to each other than they were to NYC data. It is difficult to use these fits to estimate the correct values of $\rho$ and $\tau$. Here we explore the central question of how similar SIR solutions are to each other when their parameters are quite dissimilar. Figures~\ref{fit-four-various-tau} and \ref{fit-four-various-rho} are designed to address that question.
In Figure~\ref{fit-four-various-rho}, we choose four values for 
$\rho= 1.5, 2, 3, 5$ and the rest of the parameters $I_0, A$, and $\tau$ are set equal, and we obtain four different target sets. For each target set, for each $\tau\in[3,30]$, we let the optimization function find the best values for $I_0$, $A$, and $\rho$ such that the SIR incidence $-AS'$ has the least square fit to the target set, \ie, the smallest relative error. The relative error as a function of $\tau$ is given in Figure~\ref{fit-four-various-rho}(a).  We can see that the error remains quite low even when the value of $\tau$ is significantly different from the value that generated the target dataset. We find that for each of the four $\rho$, there is a wide range of $\tau$ with a relative error of less than $2\%$.  These different curves match each other much better than they match the Omicron outbreak data. Choosing the least square fit of the Omicron data may give an illusion that one can suggest a very good estimate of the basic reproduction number and the mean duration of infectiousness.

\section{NYC COVID-19 Omicron Data}
We harvested the raw daily case data from \\ \url{https://www.nyc.gov/site/doh/covid/covid-19-data-totals.page}. This page provides a csv file from which we used the seven-day moving (centered) average of confirmed daily case counts (Column D of the csv). We selected the period from 28 November 2021 to 19 February 2022 as the time interval of interest for Omicron, amounting to 84 days in total.  The starting date was when daily cases of B1.1.529 reached 30, and the end date is when the B1.1.529 epidemic was decreasing and was being replaced by BA.2 and its sublineages. 

The NYC local website does not provide data related to cases by variant. Hence, we used the CDC data \verb+https://covid.cdc.gov/covid-data-tracker/#datatracker-home+ for this. 
The fraction of cases for each variant is reported weekly, and by geographical regions, which are composed of several states combined together. 
Region 2 consists of New York, New Jersey, Puerto Rico and the Virgin Islands. 
We report these proportions during the period in question. To fit the continuous models, i.e., differential equation models, we converted the weekly data reporting the fraction for each strain to daily data using the `makima' interpolation routine in Matlab. 
Finally, we multiplied the daily proportion history (for Omicron) by the daily case history to obtain the presumed daily counts of the 
Omicron \black variant in NYC.
These counts are given in the Supplementary Material. The dots in Figure~\ref{fig:NYC best fit} represent the data. 
\black

\begin{figure}[H]
\begin{subfigure}[c]{0.5\columnwidth}
    \centering
    \includegraphics[width=\columnwidth]{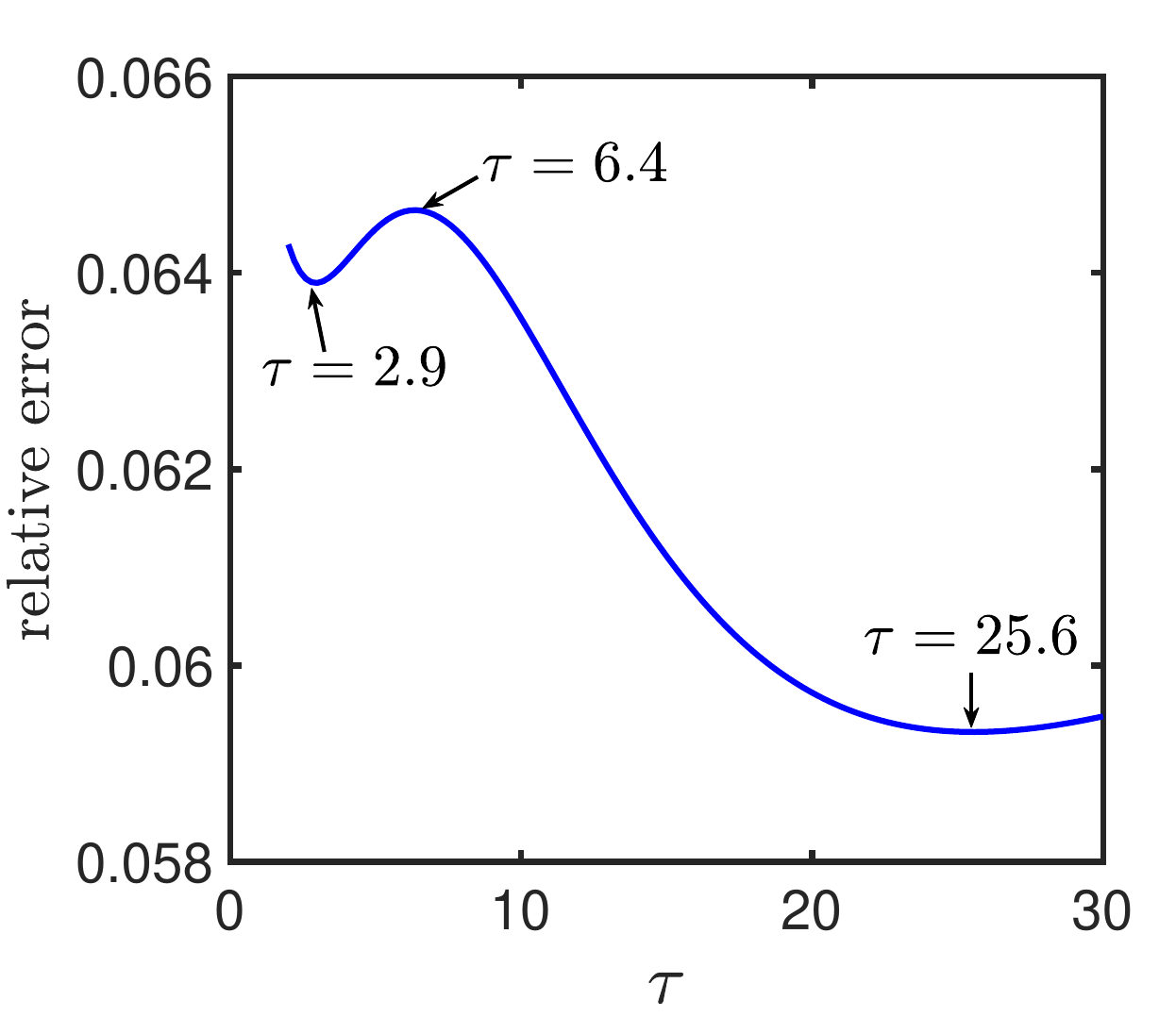}
    \caption{}
\end{subfigure}
\begin{subfigure}[c]{0.5\columnwidth}
    \centering   \includegraphics[width=1.08\columnwidth]{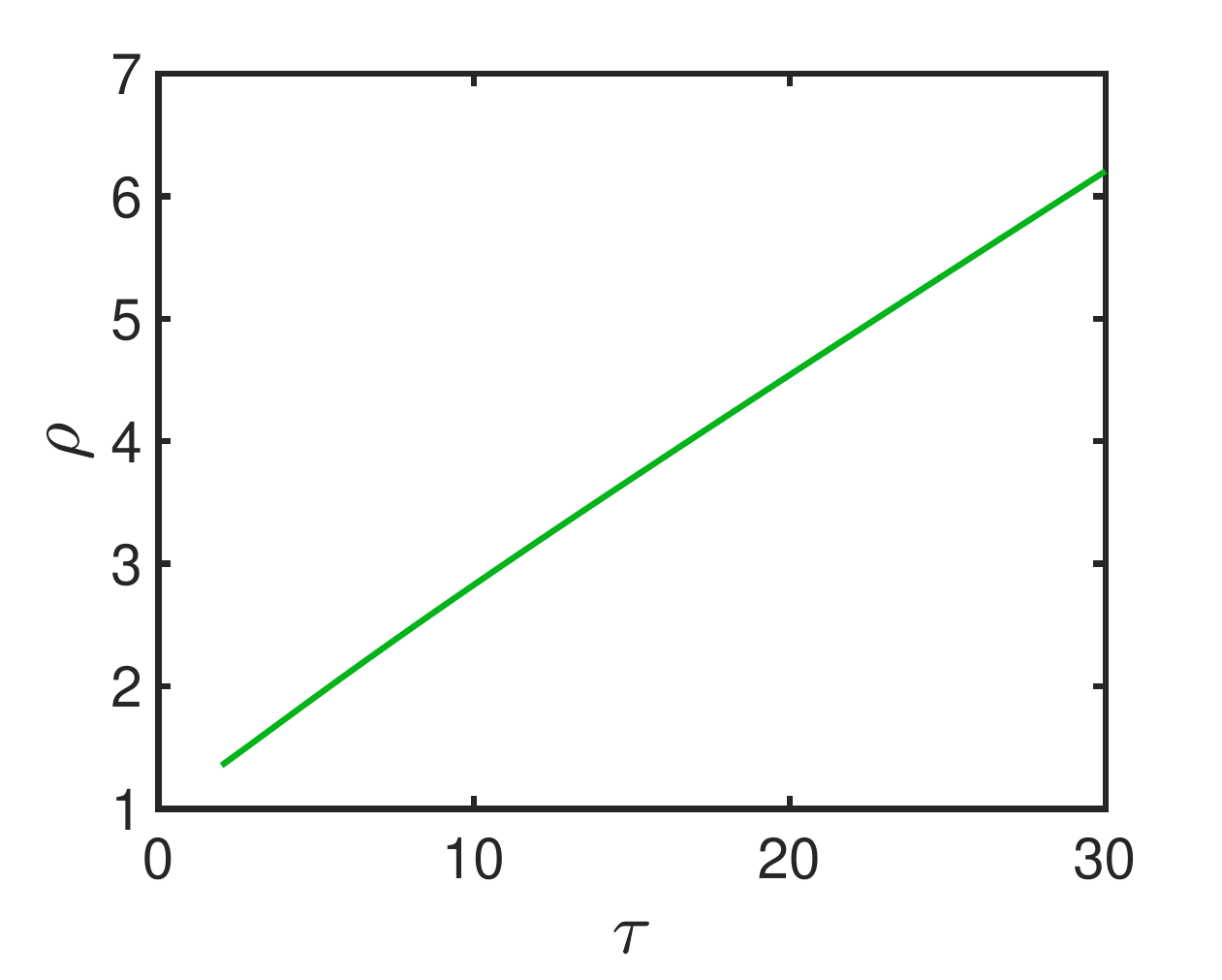}
    \caption{}
\end{subfigure}
\caption{{\bf\BF How least square fit to the NYC data varies as $\tau$ varies}. For each $\tau$ value between 2 and 30, we find the least square fit to NYC data. The corresponding relative errors are represented in panel(a). As shown in the plot, the differences in the relative errors are less than 0.005. For each $\tau$ value, the optimization function optimizes the relative error over the parameters $\rho$, $A$, and $I_0$. For each $\tau$ value, the corresponding estimated value by optimization function for $\rho$ is represented in panel(b). As shown in panel(b), there is almost a linear relationship between $\tau$ and $\rho$; as $\tau$ increases, the corresponding $\rho$ value increases.}
\label{NYC 2 parts}
\end{figure}

In Figure~\ref{fig:NYC best fit}, two least squares fits, the two solid curves, are shown. To obtain these fits, we fix the value of $\tau$ and then optimize over the rest of the parameters, $\rho, I_0$ and $A$. The exact values of the parameters are shown in Table~\ref{suptable nyc} in the supplementary material. The least squares fits are very close to each other (in terms of relative error), which means either of them can be chosen as the best fit. However, the parameter values corresponding to each fit are quite different from each other. For example, the $\tau = 2.9$ for the black curve and $\tau=25.6$ for the red curve. So it is impossible to correctly estimate the outbreak characteristics using these fits. To further illustrate the problem, we have created another simulation; we vary the value of $\tau$ from 2 to 30, and for each value of $\tau$, we optimize over the parameters $\rho$, $A$, and $I_0$. The relative error of the least squares fit from the NYC data, corresponding to each value of $\tau$ is shown in Figure~\ref{NYC 2 parts}(a). For all values of $\tau$ in the interval (2.9, 25.6), the SIR least squares fits differ from the data by about the same amount, about $6\%$, see Figure~\ref{NYC 2 parts}(a).

\section{London COVID-19 Omicron Data}
The raw data was harvested from\\
\url{https://coronavirus.data.gov.uk/details/cases?areaType=region&areaName=London}. We selected the period from 21 November 2021 to 12 February 2022 as the duration to focus on, amounting to 84 days in total. In this case, the data itself does not include seven-day averaging, so we implemented it manually. The variant proportions are available for the UK as a whole and on a weekly basis. Once again, we used the interpolation routine to convert it to a daily proportion and thereby obtained a daily case history, which is given in the Supplementary material.

The dots in Figure~\ref{fig:London fit Excluding}, which show the daily new cases of London Omicron data, have a dip in late December and early January. We believe this is a reporting problem due to the Christmas through New Year period, see Figure~\ref{supfig:London Raw}. To reflect this aberration, when we compute the least-squares fits, we have excluded from consideration the interval from day 37 to day 50, i.e., a fortnight surrounding the holiday period. Two local least squares fits to London data are shown with solid curves in Figure~\ref{fig:London fit Excluding}. To obtain each of these fits, first, we fix the value of $\tau$ and then optimize the  
relative error over the parameters $\rho, I_0$ and $A$. To see the chosen and estimated parameter values for the two least squares fits of Figure~\ref{fig:London fit Excluding}, see Table~\ref{suptable EGLL}
in the supplementary material.
The values of $\tau$ are where the local minima of relative error as a function of $\tau$ occur. To see relative error as function of $\tau$ see Figure~\ref{supfig: London tau vs RE excluding} in supplementary material. 

As it is illustrated in Figure~\ref{fig:London fit Excluding} and Figure~\ref{supfig: London tau vs RE excluding}, the weighted fractional incidence of different SIR solutions of Model~\eqref{eq:SIR} with different parameters may be very close to each other and also close to real data, which makes it difficult to use an SIR model for estimating correct parameter values, the same pattern was seen for both London and NYC Omicron data.
\begin{figure}[H]
    \centering
    \includegraphics[scale=0.32]{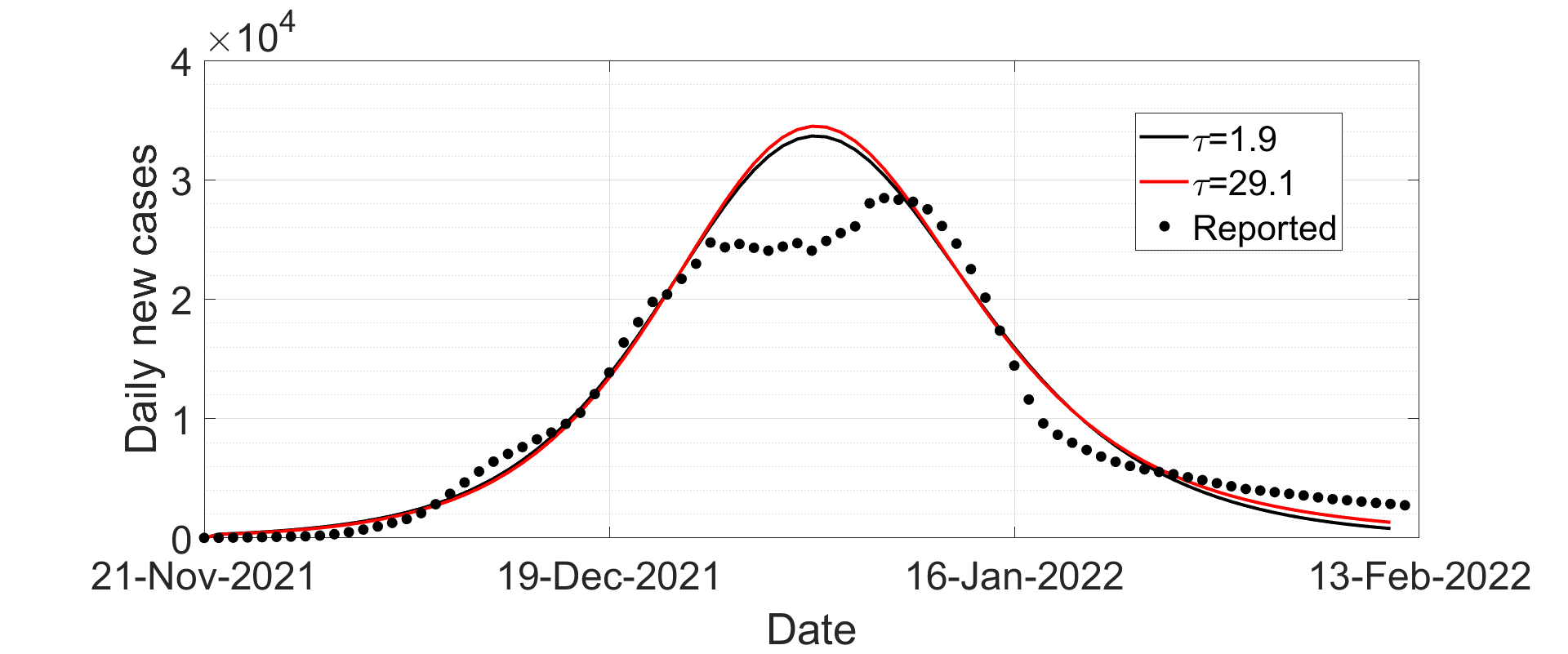}
    \caption{\textbf{Fitting SIR solutions to London data with the holiday period excluded.} Black dots denote centered 7-day average Omicron case counts In London. The black and red solid curves are two least-squares fits of Model~\eqref{eq:SIR} to London data. As the plot 
    shows the two SIR solutions are quite similar and lie close to each other, even though their corresponding parameters ($\tau, \rho, I_0, A$) are different, see Table~\ref{suptable EGLL} in supplementary material for details. The value of $\tau$ for each fit is chosen, and then the rest of the parameter values are determined using an optimization routine. 
    The overall best least-squares fit occurs for $\tau = 29.1$ for London Omicron data (\ie, the smallest relative error) with a relative error of 11.7 percent and $\tau=1.9$  is where we find a local minimum with a relative error of 12.7 percent. To see the relative error as a function of $\tau$, see Figure~\ref{supfig: London tau vs RE excluding} in the supplementary material.
    }
    \label{fig:London fit Excluding}
\end{figure}
\section{Discussion and conclusions}
Many modeling and simulation studies of physical and biological phenomena have shown that there are many combinations of parameters that can create an acceptable fit to a target data set, meaning that the fit is not unique~\cite{spear1997large, transtrum2010nonlinear}. Here, by Theorems~\ref{thm1}~and~\ref{thmN}, we prove that ``good fits'' of the SIR Model and the general compartmental SIR model to target incidence data are not unique. In fact, we show that for each choice of the basic reproduction number, one could find a good fit. This shows the downside of using fitting for estimating an outbreak's basic reproduction number or the duration of infectiousness. 

Our principle of three linear scalings elucidates why this lack of uniqueness in fitting should be anticipated. Through this principle, we show that the SIR solution has three fundamental aspects that need to be fitted to have a good fit, and there are four fitting parameters. 
The three aspects are (a) the width of the peak defined by the standard deviation,
(b) the mean time of the outbreak, and 
(c) the total number of cases. 
The four fitting parameters are (a) $\rho$, the reproduction number, (b) $\tau$, the infectious duration, (c) $N$, the total susceptible population, and (d) the initial condition at the start of the outbreak. Given a value, say $\rho$ or $\tau$, we can choose the three remaining parameters to match the three target attributes and generate what we have called a good fit.

Since so many fits are `good', we have used the relative error to determine the least square fit. For both the cities, the relative error has local minima at two values of $\tau$. In both cases, the SIR solutions for the minima $\tau$s can be seen visually to fit each other much better than they fit the target data; see Figures \ref{fig:NYC best fit} and \ref{fig:London fit Excluding}. 
Furthermore, Figure \ref{NYC 2 parts}a shows that the relative error is almost constant for NYC, remaining between 5.9\% and 6.5\% for the entire range of $\tau$. For London, the relative error ranges between 11.7\% and 12.9\% in Figure \ref{supfig: London tau vs RE excluding}.
Because of the wide range of $\tau$ values that give similar relative errors, we cannot attach to the least-squares fits any special biological significance. 

The non-uniqueness of the SIR model fit for New York City has also been reported by Jeffrey Harris \cite{jeffrey}. He has also obtained a similar fit for his target datasets, for which he does not give much credence. Our linear scalings principle explains the existence of good fits for every value of $\rho$. The good fits and least-squares fits for each $\rho$ are quite similar to each other.

In Figures~\ref{fit-four-various-rho}~and~\ref{NYC 2 parts} we can see that the least-squares fit $\rho$ and $\tau$ lie almost on a straight line. This can be understood by calculating the exponential growth rate at the start of the outbreak, 
when $S\approx 1$. From Equation~\eqref{eq:SIR}, we obtain:
$$
\frac{1}{I}\frac{dI}{dt} =  \frac{\rho -1}{\tau}.
$$

Hence all fits to a target should all have approximately 
the same exponential growth rate (EGR) $=\frac{\rho-1}{\tau}$, which is a constant that depends on the choice of target outbreak. We observed the same pattern in the London data, so we only show the NYC graphs. 

In both Figures~\ref{fig:NYC best fit}~and~\ref{fig:London fit Excluding}, we can see that the model solution lies above the data initially and below the target data towards the end of the period being fitted. This may be due to a structural weakness of the SIR model in describing the beginnings and ends of outbreaks, or it may represent errors in data collection when most COVID cases are not Omicron.

Of course, transmission parameters like $\rho$ and $\tau$ can be and are published, and for COVID-19, these may differ for different strains. These estimates are determined from extra data. We simply show that such estimates cannot be obtained purely from the incidence data.

The three linear scalings of $\tau, \mu,$ and $A$ show that a good fit exists for every value of $\rho$, and the exponential growth rate tells us that for each $\rho$, there is a corresponding $\tau$, (Figure~\ref{NYC 2 parts}. 
Now suppose we measure either one of these through direct experiment ($\rho$ through contact tracing and $\tau$ through case-tracing of individual cases and their transmissions). Then, the degeneracy will be broken, and after that, if we obtain a least-squares fit, then that will have significant credibility. 

For a more complicated model with more transmission parameters, i.e., higher-dimensional versions of $\rho$ and $\tau$, the degeneracy problem is much worse. Our Theorem~\ref{thmN} shows, for example, that if there are $N=30$ compartments in our higher-dimensional SIR model \eqref{eq:SIR-N}, there is a 900-dimensional space of parameter sets for which there is a good fit. Hence, the overfitting phenomenon we are identifying is made worse by adding compartments.   

In conclusion, we have shown that the SIR model has mathematical properties that epidemiologists should know about and might wish to consider when fitting SIR solutions to data. 

\nolinenumbers
\section*{Acknowledgements}
We thank Dr.~Ellen D Yorke, Dr.~Abba Gumel, Chirag Adwani, and Binod Pant  for valuable discussions and comments on the project.

\bibliographystyle{unsrt}
\bibliography{Ref}
 \appendix
\counterwithin{figure}{section}
\counterwithin{table}{section}
\section*{Supplementary Material}\label{sec:supplement}
\section{Tables}\label{app:tables}
\begin{table}[H]
    \centering
    \begin{tabular}{|c|c|c|c|c|c|c|}
         \hline
      $\tau$ & $\rho$ & $A$ & $I_{0}$ &$S_{0}$& relative error & Infected percent\\   \hline
      2 & 1.2693  & 2.004 & 0.0003 &0.9986 &0.0306& 39.20\\
         \hline
           7 &  2 & 1.0 & 0.001 &  0.9980& --& 79.68\\
         \hline
     10 &  2.4285 & 0.9082 & 0.0013 & 0.9978&0.007& 88.28\\ \hline
     20 & 3.7818  & 0.8269 & 0.0021 &0.9972 & 0.0117
& 97.50\\ \hline
       \end{tabular}
       \caption{Parameter values corresponding to the four solutions in Figure~\ref{fit-four-various-tau}
       . The target has $\rho = 2$ and $\tau = 7$.
       }
        \label{suptable:SIR-vs-SIR}
\end{table}  
\begin{table}[H]
    \centering
    \begin{tabular}{|c|c|c|c|c|c|}
         \hline
      $\tau$ & $\rho$ & $A$ & $I_{0}$ & Relative error & Infected percent\\     \hline   2.9 &  1.505 & 1{,}744{,}364 &0.0003& 0.064 & 58.6 \\
      \hline
      25.6 & 5.483 & 1{,}041{,}895  & 0.0013 & 0.059 & 99.6\\ 
     \hline
     \end{tabular}
    \caption{Estimated parameter values for SIR solution yield local best fit for NYC.}
    \label{suptable nyc}
\end{table}
\begin{table}[H]
    \centering
    \begin{tabular}{|c|c|c|c|c|c|c|}
         \hline
      $\tau$ & $\rho$ & $A$ & $I_{0}$ & Relative error & Infected percent\\   
      \hline
     1.9 &  1.280 & 2{,}421{,}000 & 0.00016 & 12.7 & 40.3\\
      \hline
     29.1 & 5.344 & 1{,}009{,}000 &  0.0013 & 11.7 & 82.3\\ 
     \hline
     \end{tabular}
    \caption{Estimated parameter values for SIR solution yield local best fit for London data with holiday period excluded.}
    \label{suptable EGLL}
\end{table}
\section{Figures}
\begin{figure}[H]
    \centering
    \includegraphics[width=\linewidth]{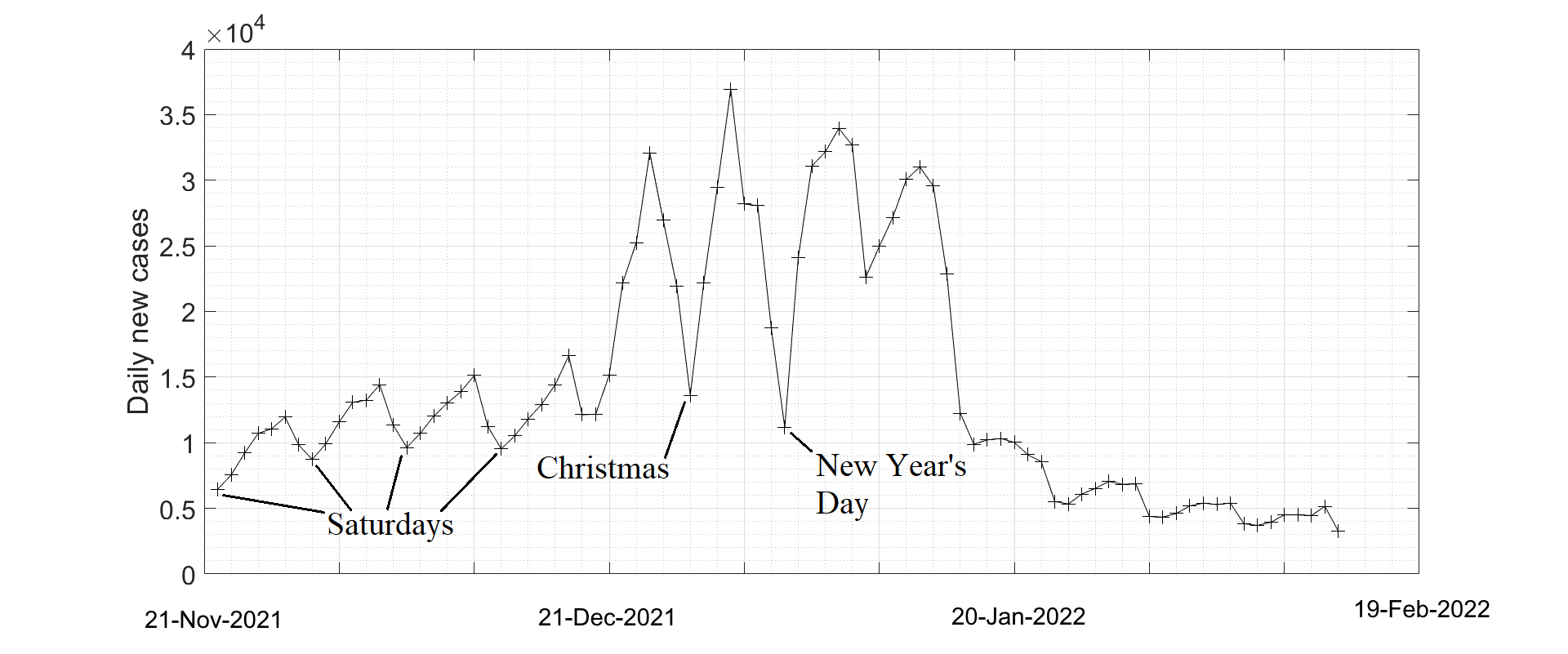}
    \caption{The daily raw data of all COVID-19 cases for London.}
    \label{supfig:London Raw}
\end{figure}
\begin{figure}[H]
    \centering
    \includegraphics[width=\linewidth]{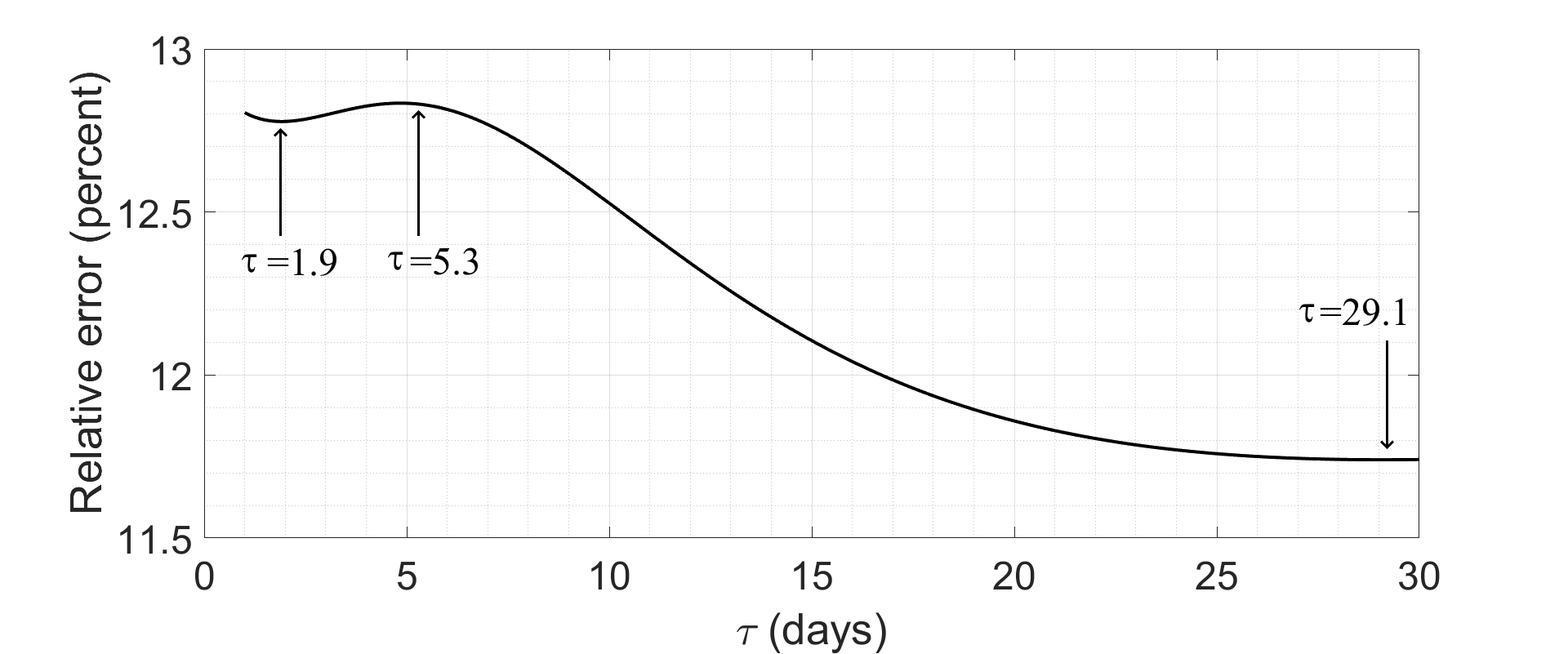}
    \caption{\textbf{Fit of the London data using SIR model with the central period excluded.}     The relative error has two local minima at $\tau=1.9$ and $\tau=29.1$ days. Further details of these fits are in Table~\ref{suptable EGLL}.}
    \label{supfig: London tau vs RE excluding}
\end{figure}
\section{Data}\label{supappendix data}
Reported Omicron cases (the sum of the numbers used in this study):

NYC: 1,024,164

London: 906,656

The variant proportion of Omicron in US HHS Region 2 over the twelve-week period in consideration is as follows :
\begin{equation} \nonumber
\begin{array}{cccccccccccc}
   2.1 & 19.2 & 55.7 & 84.7 & 90.1 & 95.7 & 97.9 & 98.7 & 97.9 & 98.2 \\
\end{array}
\end{equation}

The variant proportion of Omicron in UK over the twelve-week period in consideration is as follows :
\begin{equation}\nonumber
    \begin{array}{cccccccccccc}
         0.2 & 1.4 & 13.2 & 58.0 & 82.3 & 94.6 & 94.1 & 97.4 & 96.2 & 92.8 & 85.4 & 76.5 \\ 
    \end{array}
\end{equation}

NYC's moving (centered) seven-day average of daily cases of the Omicron variant for 84 consecutive days beginning Nov. 28, 2021. The cases of each date are averaged with the three previous days and the following three days. This data is shown with ten consecutive days per row:\\
\begin{equation}\nonumber
    \begin{array}{rrrrrrrrrr}
    32 & 53 & 87 & 135 & 211 & 290 & 375 & 470 & 589 & 755 \\ 
    955 & 1178 & 1442 & 1673 & 1968 & 2597 & 3666 & 5083 & 6747 & 8447 \\
    9855 & 11158 & 14036 & 16818 & 19645 & 22252 & 22963 & 22814 & 24763 & 28328 \\ 
    31161 & 34062 & 36696 & 38696 & 39641 & 41217 & 41945 & 42267 & 40814 & 39136 \\ 
    38320 & 39005 & 36322 & 31771 & 27050 & 23724 & 20648 & 19137 & 17651 & 16448 \\ 
    14047 & 12861 & 11354 & 9932 & 8737 & 8250 & 7708 & 7102 & 6140 & 5391 \\ 
    4873 & 4398 & 3900 & 3542 & 3263 & 2850 & 2490 & 2222 & 1987 & 1816 \\ 
    1869 & 1785 & 1570 & 1462 & 1317 & 1203 & 1119 & 1069 & 1005 & 918 \\ 
    832 & 772 & 625 & 659 & & & & &\\
    \end{array}
\end{equation}

London's moving seven-day average of daily cases daily cases of the Omicron variant for 84 consecutive days beginning Nov. 21, 2021. This data is shown with ten consecutive days per row:\\

\begin{equation}\nonumber
    \begin{array}{rrrrrrrrrr}
    17 & 23 & 36 & 53 & 74 & 98 & 125 & 152 & 209 & 324\\
    493 & 709 & 970 & 1268 & 1591 & 2089 & 2833 & 3709 & 4653 & 5575 \\
    6397 & 7047 & 7623 & 8269 & 8838 & 9563 & 10492 & 12060 & 13868 & 16374 \\
    18086 & 19776 & 20400 & 21705 & 22977 & 24750 & 24351 & 24626 & 24306 & 24067 \\
    24410 & 24694 & 24067 & 24884 & 25545 & 26101 & 28034 & 28471 & 28331 & 28162 \\
    27532 & 26132 & 24654 & 22522 & 20133 & 17371 & 14445 & 11599 & 9604 & 8645 \\
    7983 & 7382 & 6826 & 6384 & 6036 & 5754 & 5543 & 5346 & 5089 & 4854 \\
    4588 & 4338 & 4104 & 3978 & 3843 & 3706 & 3567 & 3407 & 3256 & 3168 \\
    3049 & 2942 & 2858 & 2743\\
    \end{array}
\end{equation}

\end{document}